\documentclass[useAMS, usenatbib, onecolumn]{mn2e}
\usepackage{graphicx}
\usepackage{amssymb}

\newcommand\be{\begin{equation}}
\newcommand\ee{\end{equation}}
\newcommand\ba{\begin{eqnarray}}
\newcommand\ea{\end{eqnarray}}

\newcommand\bb[1]{\mbox{\boldmath{$#1$}}}
\newcommand{\Alfven}{Alfv\'{e}n~}
\newcommand\tomega{{\tilde\omega}}

\def\go{\mathrel{\raise.3ex\hbox{$>$}\mkern-14mu
             \lower0.6ex\hbox{$\sim$}}}
\def\lo{\mathrel{\raise.3ex\hbox{$<$}\mkern-14mu
             \lower0.6ex\hbox{$\sim$}}}

\title[Dynamics of the innermost accretion flows around compact
  objects] {Dynamics of the Innermost Accretion Flows Around Compact
  Objects: Magnetosphere-Disc Interface, Global Oscillations and
  Instabilities}
\author[W.~Fu \& D.~Lai]
{Wen Fu$^{1,2}$\thanks{Email: wenfu@astro.cornell.edu (WF);
dong@astro.cornell.edu (DL)} and Dong Lai$^{1}$\footnotemark[1]\\
$^1$Department of Astronomy, Cornell University, Ithaca, NY 14853, USA\\
$^2$Theoretical Division, Los Alamos National Laboratory, Los Alamos, NM 87545, USA}

\pagerange{\pageref{firstpage}--\pageref{lastpage}} \pubyear{2012}

\begin{document}

\label{firstpage}
\maketitle

\begin{abstract}
We study global non-axisymmetric oscillation modes and instabilities
in magnetosphere-disc systems, as expected in neutron star X-ray binaries and possibly also in accreting
black hole systems. Our two-dimensional magnetosphere-disc model
consists of a Keplerian disc in contact with an uniformly rotating
magnetosphere with low plasma density. Two types of global overstable modes exist
in such systems, the interface modes and the disc inertial-acoustic
modes. We examine various physical
effects and parameters that influence the properties of these
oscillation modes, particularly their growth rates, including the
magnetosphere field configuration, the velocity and density contrasts
across the magnetosphere-disc interface, the rotation profile
(with Newtonian or pseudo General Relativistic potential), the sound speed and magnetic field of
the disc.  The interface modes are driven unstable by Rayleigh-Taylor
and Kelvin-Helmholtz instabilities, but can be stabilized by the
toroidal field (through magnetic tension) and disc differential
rotation (through finite vorticity). General relativity increases
their growth rates by modifying the disc vorticity outside the
magnetosphere boundary. The interface modes may also be affected by
wave absorption associated with corotation resonance in the disc.  In
the presence of a magnetosphere, the inertial-acoustic modes are effectively trapped at the innermost region of
the relativistic disc just outside the interface. They are driven unstable by wave
absorption at the corotation resonance, but can be stabilized by modest
disc magnetic fields. The overstable oscillation modes studied in this
paper have characteristic properties that make them possible candidates
for the quasi-periodic oscillations observed in X-ray
binaries.
\end{abstract}


\section{Introduction}

Quasi-periodic oscillations (QPOs) in X-ray fluxes, with frequencies
comparable to the dynamical frequencies of neutron stars (NSs) and
black holes (BHs), have been observed since the 1990s in many X-ray
binary systems (e.g., van der Klis 2006), largely thanks to NASA's
{\it Rossi X-ray Timing Explorer} (Swank 1999). They are of great
interest because they probe the dynamics of the innermost accretion
follows around NSs and BHs, and may potentially help constrain the
physics of dense nuclear matter and strong gravity.

The phenomenology of kHz QPOs in NS low-mass X-ray binaries is well
established (van der Klis 2006). They usually occur in pairs, with
frequencies between 300~Hz and 1.2~kHz, both varying significantly as
a function of the X-ray flux. There is some correlation between the
upper frequency $\nu_u$ and the lower frequency $\nu_l$.  Spectral
analysis indicates most kHz QPOs arise from variations in the inner
disk boundary layers (Gilfanov et al.~2003).  High-frequency QPOs
(HFQPOs) (40-450~Hz) in BH X-ray binaries have also been intensively
studied (Remillard \& McClintock 2006; Belloni et al.~2011; Altamirano, Belloni \&
Linares et al.~2011; Altamirano
\& Belloni 2012),
although the signals are much weaker and transient. They are only
observed in the ``transitional state'' (or ``steep power law state'')
of the X-ray binaries, and have low amplitudes 
and low coherence. 
Their frequencies do not vary significantly in response to sizeable
(factors of 3-4) luminosity changes. Several systems show pairs of
QPOs with frequency ratios close to $2:3$ (Abramowicz \& Kluzniak 2001), 
$5:3$ (Kluzniak \& Abramowicz 2002), and $5:2$ (Rebusco, Moskalik \& 
Kluzniak 2012).

Despite much observational progress, the physical origin of HFQPOs
remains elusive. Based on the general consensus that HFQPOs are
associated with the dynamics of innermost accretion flows, a number of
ideas/models with various degrees of sophistication have been proposed
or studied. The general notions of hot blobs exhibiting (test-mass)
orbital motion and relativistic precession (Stella et al.~1999),
nonlinear resonances (Abramowicz \& Kluzniak 2001), and in the case
of NS systems, beats between the orbital frequency and stellar spin
(Miller et al.~1998), have been the most popular among the observers,
although it is not clear how these generic ideas are realized in
real fluid dynamical models of accretion flows. Other ideas for NS kHz
QPOs are discussed in Bachetti et al.~(2010).
In the case of HFQPOs in BH X-ray binaries, the types of models that go
beyond the ``test-mass motion'' idea are those based on global oscillations
of accretion flows, including discs, tori and boundary layers [see
below; see also Sect.~1 of Lai \& Tsang (2009) for a critical review].
Systematic studies of accretion flow dynamics, combining semi-analytic
calculations (for idealized systems to extract physical
insights/ingredients) and full numerical simulations (for systems with
increasing sophistications and realisms) are essential for making
further progress in this field.

In this paper we study a class of models involving global oscillations
of inner accretion discs and magnetosphere boundary layers around NSs
or BHs. Our model setup (see Fig.~1) consists of an uniformly rotating
magnetosphere with both toroidal and poloidal magnetic fields and low
plasma density, surrounded by a geometrically thin (Keplerian) disc
with high fluid density and zero (or low) magnetic field.  To make 
semi-analytical linear perturbation analysis feasible, our model is
two-dimensional with no vertical dependence (i.e., it is height
averaged). We consider both Newtonian discs and general relativistic
(GR) discs (modelled by the pseudo-Newtonian potential) -- we will see
that GR can qualitatively affect the linear oscillation modes of the
system by changing the disc vorticity profile.  
This kind of magnetosphere-disc systems can form in accreting NS
systems when the NS magnetic field holds off the accretion disc
at a certain distance from the stellar surface (e.g., Ghosh \& Lamb
1978). They may also be applied to accreting BH systems when
magnetic fields advect inwards in the accretion disc and accumulate
near the inner edge of the disc (e.g. Bisnovatyi-Kogan \& Ruzmaikin
1974, 1976; Igumenshchev, Narayan \& Abramowicz 2003; Rothstein \&
Lovelace 2008, McKinney, Tchekhovskoy \& Blandford 2012). It is usually thought that during the
``transitional state'' (when HFQPOs are observed) of BH X-ray
binaries, the accretion flow consists of a thin, thermal disc,
truncated by a hot, tenuous (and perhaps highly magnetized) inner
corona (e.g., Done et al.~2007; Oda et al.~2010).
While the true nature of this flow is unclear, our simple
magnetosphere-disc setup may represent an idealization of 
such a flow.

\begin{figure}
\begin{center}
\includegraphics[width=0.75\textwidth]{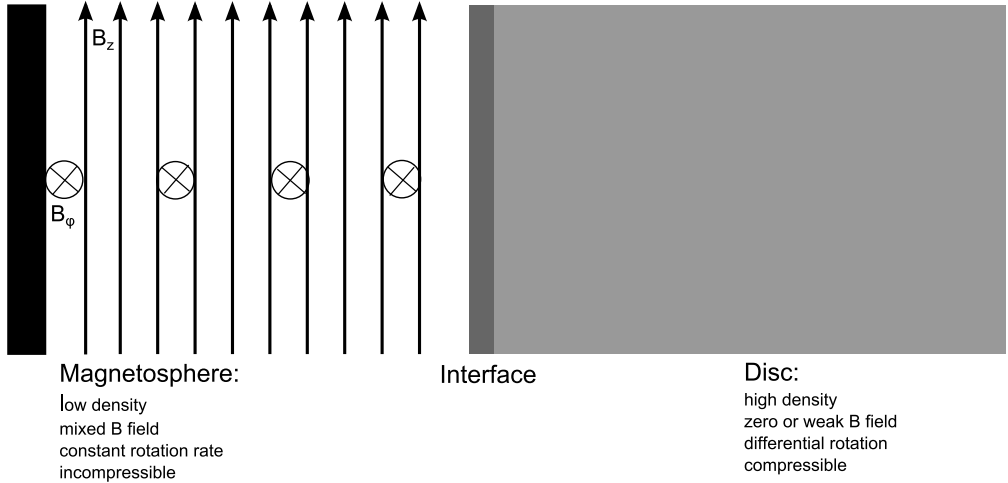}
\caption{Schematic description of the cylindrical magnetosphere-disc
  model considered in this paper. The thick black line on the left
  represents the central compact object. Right next to it is a region
  of low density plasma threaded by both vertical and toroidal
  magnetic fields. The disc (on the right) consists of fluid of high
  density (compared to the magnetosphere) and zero or weak magnetic
  field. These two regions are separated by a thin interface/boundary
  layer.}
\label{fig:interface}
\end{center}
\end{figure}

Loosely speaking, two types of global oscillation modes exist in our
2D magnetosphere-disc systems: {\it inertial-acoustic modes of the disc} and
{\it interface modes}. Both types of oscillations can be overstable, but
driven by different physical mechanisms.
Our work represents an extension of related previous studies,
which have focused on disc oscillations or interface oscillations separately.

{\it (i) Disc Oscillations:} 
Although relativistic fluid discs can support 
different types of oscillation modes (Okazaki, Kato \& Fukue 1987;
Nowak \& Wagoner 1991; see Wagoner 1999, Kato 2001 and Kato et al.~2008
for reviews),
our focus in this paper, as well in our previous papers 
(Lai \& Tsang 2009; Tsang \& Lai 2008,~2009c; Fu \& Lai 2011a),
is on the inertial-acoustic modes (also called p-modes). These modes 
have no vertical structure (i.e., their wave functions, such as the pressure 
perturbation, have no node in the vertical direction), 
and are adequately captured by our 2D model. 

On the other hand, our magnetosphere-disc model does not capture 
other discoseismic modes, including g-modes (also called inertial-gravity 
modes), c-modes and vertical p-modes 
(see Wagoner 1999; Kato et al.~2008; Kato 2011a,~b).
All these modes have vertical structures.
Non-axisymmetric g-modes and c-modes are either damped due to
corotation resonance (Kato 2003a, Li, Goodman \& Narayan 2003; Tsang
\& Lai 2009a) or their self-trapping zones are easily ``destroyed'' by
disc magnetic fields (Fu \& Lai 2009). Although axisymmetric g-modes
($m=0$) can be resonantly excited by global disc deformations through
nonlinear effects (Kato 2003b,~2008; Ferreira \& Ogilvie 2008), their
trapping zones disappear even for a weak (sub-thermal) disc magnetic
field.  By contrast, the basic wave properties (e.g., propagation
diagram) of p-modes are not strongly affected by disc magnetic fields
(Fu \& Lai 2009), and it is likely that these p-modes are robust in
the presence of disc turbulence (see Arras, Blaes \& Turner 2006;
Reynolds \& Miller 2009). 

Our previous papers (Lai \& Tsang 2009; Tsang \& Lai
2008,~2009c; Fu \& Lai 2011a) have shown that non-axisymmetric p-modes
trapped in the inner-most region of a BH accretion disc are a
promising candidate for explaining HFQPOs in BH x-ray binaries because
their frequencies naturally match the observed values without
fine-tuning of the disc properties, and because they become overstable
due to wave absorption at corotation point when the general relativistic 
effect on the disc rotation profile is taken into account. Although a toroidal
disc magnetic field tends to suppress the instability (Fu \& Lai 2011a),
a large-scale poloidal field may enhance the instability (Tagger \&
Pallet 1999; Tagger \& Varniere 2006).  Most of these
previous papers adopt a reflective boundary condition at the inner
edge of the disc -- this is important for the trapping of disc p-modes
(see Lai \& Tsang 2009). How such a reflection can be achieved was not
clear. We will show in this paper that the disc-magnetosphere boundary
serves as a ``reflector'' for the disc p-modes.

{\it (ii) Interface Oscillations:}
Li \& Narayan (2004) were the first to consider the interface modes
between a magnetosphere (with a vertical magnetic field) and an 
incompressible disc in the cylindrical approximation (i.e., no 
$z$-dependence). They showed that the interface modes can be subject to 
Rayleigh-Taylor instability and/or Kelvin-Helmholtz instability, 
depending on the density contrast and velocity shear across the interface.
The mode frequencies roughly scale as $m\Omega_{\rm in}$,
where $m=1,\,2,\,3\ldots$ are the azimuthal mode number and 
$\Omega_{\rm in}$ is the angular frequency of the disc flow at the
interface, making the interface modes a viable candidate for 
explaining HFQPOs when the interface radius $r_{\rm in}$ 
is suitably adjusted.
Tsang \& Lai (2009b) generalized the analysis of Li \& Narayan (2004)
by considering compressible discs. They showed that 
a relatively large disc sound speed is necessary to overcome the
stabilizing effect of disc differential rotation and thereby 
maintain the mode growth. Besides linear
calculations, numerical simulations of the 
magnetosphere-disc interface have also been presented in a number of
papers (see Kulkarni \& Romannova 2008; Romanova, Kulkarni \&
Lovelace 2008 and references therein).

One ingredient that has been missing from both Li \& Narayan (2004)
and Tsang \& Lai (2009b) is toroidal magnetic fields in the
magnetosphere, which could be a very important component (Ghosh \&
Lamb 1978; Ikhsanov \& Pustil'nik 1996).  In this paper, we generalize
previous calculations of global non-axisymmetric modes confined near
the interface of the magnetosphere-disc system by taking into account
toroidal magnetic fields in the magnetosphere.  We aim to examine
the possibility of large-scale Rayleigh-Taylor/Kelvin-Helmholtz 
instabilities of the interface in the presence of magnetic 
fields and disc differential rotation. 

Overall, the goal of this paper is to present a complete analysis 
of large-scale non-axisymmetric modes (including their instabilities)
in the magnetosphere boundary and in the surrounding disc, and to assess
their viabilities for explaining HFQPOs in accreting compact binaries.

We note that in addition to Li \& Narayan (2004) and Tsang \& Lai (2009b), our
study on the interface modes complements several other works on the
instability of magnetized accretion discs.  For example, Lubow \&
Spruit (1995) and Spruit, Stehle \& Papaloizou (1995) considered the
magnetic interchange instability of a thin rotating disc when a
poloidal magnetic field provides some radial support in the disc.
Lovelace, Turner \& Romanova (2009) studied the instability of a
magnetopause for cases where the shear layer has appreciable radial width
and found that the Rossby wave instability may arise in the shear layer.
Lovelace, Romanova \& Newman (2010) considered a setup similar to
the present paper, but they included only vertical field in the magnetosphere,
and focused on small-scale (with radial wavelength $\ll r$) Kelvin-Helmholtz
modes in the boundary layer.

This paper is the fourth in our series of papers on the effects of
magnetic fields on the global instabilities of various astrophysical
flows, with the previous three focusing on black hole accretion discs
(Fu \& Lai 2011a), accretion tori (Fu \& Lai 2011b) and rotating
proto-neutron stars (Fu \& Lai 2011c), respectively.  Our paper is
organized as follows. In Section 2, we present the setup of our
cylindrical magnetosphere-disc model and the dynamical equations for
wave modes in the model.  Section 3 focuses on the interface modes
when Newtonian potential is used for disc rotation, while Section 4
deals with both the interface modes and disc p-modes when the GR
effect (using pseudo-Newtonian potential) is included in the disc
rotation. We discuss our results and conclude in Section 5.

\section{Equilibrium and Perturbation equations}

We consider a magnetosphere-disc model similar to the one in Tsang \&
Lai (2009b). It consists of a magnetosphere region where magnetic
pressure dominates over gas pressure and a disc region which has high
density compared to the magnetosphere (see Fig.~\ref{fig:interface}).
These two regions are separated by an interface (boundary
layer). Unlike Tsang \& Lai (2009b), who considered only vertical
magnetic field for the magnetosphere, we take into account both
vertical and toroidal fields.  In the disc region, gas pressure dominates over magnetic pressure. Since any initial
poloidal field is likely to generate a dominating toroidal field
due to the disc differential rotation, for simplicity we take the disc
to be threaded by toroidal B field only. We assume that flows in both
regions are non-self-gravitating, satisfying the usual ideal MHD
equations
\be
{\partial{\rho} \over \partial t}
+\nabla\cdot(\rho \bb{v})=0,
\label{eq:mhd1}
\ee
\be
{\partial{\bb{v}} \over \partial
t}+({\bb{v}}\cdot\nabla){\bb{v}}
=-\frac{1}{\rho}\nabla \Pi-{\nabla \Phi}
+\frac{1}{4\pi\rho}(\bb{B}\cdot\nabla)\bb{B},
\label{eq:mhd2}
\ee
\be
{{\partial {\bb{B}}} \over \partial t}=\nabla\times({\bb{v}}
\times{\bb{B}}),
\label{eq:mhd3}
\ee
where $\rho,\,P,\,\bb{v}$ are the fluid density, pressure and velocity,
$\Phi$ is the gravitational potential due to the central compact object, and
\be
\Pi \equiv P+\frac{\bb{B}^2}{8\pi}
\label{eq:tp}
\ee
is the total pressure.  The magnetic field $\bb{B}$ also satisfies the
equation $\nabla\cdot\bb{B}=0$.

We adopt the cylindrical coordinates $(r, \phi, z)$ which are centered
on the central object and have the $z$-axis in the direction
perpendicular to the disc plane.  The unperturbed background disc has
a velocity field $\bb{v}=r\Omega(r)\bb{\hat \phi}$, and the magnetic
field may consist of both toroidal and vertical components $\bb{B}=
B_\phi(r)\bb{\hat \phi}+B_z(r) \bb{\hat z}$. The gravitational
acceleration in radial direction is defined as
\be
g=\frac{d\Phi}{dr}
\ee
so that $-\nabla\Phi=-g{\bb{\hat r}}$ and $g=r\Omega_{\rm K}^2>0$, where
$\Omega_{\rm K}$ is the angular frequency for a test mass (the Keplerian
frequency).
Thus the radial force balance equation reads
\be
\rho g=\rho r\Omega^2-\frac{d\Pi}{dr}-\frac{B_{\phi}^2}{4\pi r}.
\label{eq:fb}
\ee

To investigate the dynamical properties of the flow, we perturb the
MHD Eqs.~(\ref{eq:mhd1})-(\ref{eq:mhd3}) by rewriting any physical
variable $f$ as $f+\delta f$ with $|\delta f| \ll |f|$. Since the
unperturbed state is axisymmetric and steady, we consider 
all perturbation variables having the form
$\delta f \propto {\rm e}^{{\rm i}m\phi-{\rm i}\omega t}$, with
$m$ being the azimuthal mode number and $\omega$ the wave
frequency. Note that the background flow and magnetic field have 
no $z$-dependence and we assume that the perturbations
also have no $z$-dependence.
The resulting linearized perturbation equations 
contain the variables $\delta \bb{v}$, $\delta \rho$, $\delta P$, $\delta
\Pi$ and $\delta \bb{B}$. For mathematical convenience, we define a
new variable
\be
\delta h=\frac{\delta \Pi}{\rho}=
\frac{\delta P}{\rho}+\frac{\bb{B}\cdot \delta \bb{B}}{4\pi\rho}.
\ee
Moreover, using $\Delta \bb{v}=\delta
\bb{v}+\bb{\xi}\cdot\nabla\bb{v}=d\bb{\xi}/dt
=-i\omega\bb{\xi}+(\bb{v}\cdot \nabla)\bb{\xi}$, we find that the
Eulerian perturbation $\delta \bb{v}$ is related to the Lagrangian
displacement vector $\bb{\xi}$ by $\delta
\bb{v}=-i\tomega\bb{\xi}-r\Omega'\xi_r \bb{\hat{\phi}}$, where prime denotes
radial derivative and 
\be
\tomega=\omega-m\Omega,
\ee
is the Doppler-shifted wave frequency. In the next two subsections, we
will combine the perturbations equations into two first-order
differential equations (ODEs) in terms of $\xi_r$ and $\delta h$ for the
magnetosphere region and the disc region, respectively.

\subsection{The magnetosphere}

In the magnetosphere region ($r < r_{\rm in}$),
the flow is assumed to be incompressible and have uniform density
($\rho$ is constant). For this particular magnetized fluid system, the
detailed linearized perturbation equations have been given in Fu \&
Lai (2011b). Here we just display the final two ODEs for $\xi_r$ and
$\delta h$:
\be
\frac{d\xi_r}{dr}=A_{11}\xi_r+A_{12}\delta h,
\label{eq:odem1}
\ee
\be
\frac{d\delta h}{dr}=A_{21}\xi_r+A_{22}\delta h,
\label{eq:odem2}
\ee
where
\be
A_{11}=-\frac{1}{r}\frac{\tomega^2-2m\tomega\Omega+m^2\omega_{A\phi}^2}
{\tomega^2-m^2\omega_{A\phi}^2},
\ee
\be
A_{12}=\frac{m^2}{r^2},
\ee
\be
A_{21}=\tomega^2-m^2\omega_{A\phi}^2-2r\Omega\frac{d\Omega}{dr}
+\left(2\frac{d\ln B_{\phi}}{d\ln r}-1\right)
\omega_{A\phi}^2-4\frac{(\tomega\Omega+m\omega_{A\phi}^2)^2}{(\tomega^2-m^2\omega_{A\phi}^2)},
\ee
\be
A_{22}=\frac{2m}{r}\frac{\tomega\Omega+m\omega_{A\phi}^2}{\tomega^2-m^2\omega_{A\phi}^2},
\ee
and $\omega_{A\phi}\equiv v_{A\phi}/r=B_{\phi}/({r\sqrt{4\pi\rho}})$ is
the toroidal \Alfven frequency. Note that although we assume low
plasma density for the magnetosphere, we still require that
density is not too low in order to keep \Alfven speed far less than
the speed of light. Otherwise, our original MHD equations break down
(e.g., Lovelace, Romanova \& Newman 2010). Also note that Eqs.
(\ref{eq:odem1}) and (\ref{eq:odem2}) are the same as Eqs.~(119) and
(120) (derived for a purely toroidal magnetic field) of \S 83 
in Chandrasekhar (1961). The vertical magnetic field $B_z$ does not appear 
in our perturbation equations because we assumed $k_z=0$, i.e., vertical
field lines are not perturbed.

Defining $\sigma^2=\tomega^2-m^2\omega_{A\phi}^2$, we can further
combine Eqs.~(\ref{eq:odem1}) and (\ref{eq:odem2}) into one single
equation
\be
\xi_r''+\frac{d}{dr}\ln (r^3\sigma^2) \xi_r'+\frac{1-m^2}{r^2}\xi_r=0.
\label{eq:odem3}
\ee
Two special cases are of interest:

(i) For systems with $B_\phi=0$, Eq.~(\ref{eq:odem3}) reduces to
\be
\xi_r''+(\frac{3}{r}-\frac{2m\Omega'}{\tomega})\xi_r'+\frac{1-m^2}{r^2}\xi_r=0,
\label{eq:odexi}
\ee
or equivalently
\be
W''+\frac{W'}{r}-\frac{m^2}{r^2}\left[1-\frac{r}{m\tomega}\frac{d}{dr}
\left(\frac{\kappa^2}{2\Omega}\right)\right]W=0,
\label{eq:odew}
\ee
where $W=r\delta v_r$ and 
\be 
\kappa=\left[\frac{2\Omega}{r}\frac{d}{dr}(r^2\Omega)\right]^{1/2}
\ee
is the radial epicyclic frequency. Note that Eq.~(\ref{eq:odew})
recovers Eq.~(12) in Tsang \& Lai (2009b). For either uniform rotation
profile [$\Omega={\rm const}$, $\kappa=2\Omega$, thus
$(\kappa^2/2\Omega)'=0$] or uniform angular momentum profile ($\Omega
\propto r^{-2}$, thus $\kappa=0$), it has the same simple solution
\be
W \propto r^m~~ \mbox{or}~~\delta v_r \propto r^{m-1}.
\ee
The corresponding solution for $\xi_r$ is 
\be
\xi_r \propto r^{m-1}~~\mbox{if}~~\Omega=\mbox{const;}~~
\xi_r \propto \frac{r^{m-1}}{\omega-m\Omega(r)}~~\mbox{if}~~\Omega \propto r^{-2}.
\ee

(ii) In the special case of $\Omega=\mbox{const}$ and $B_{\phi}
\propto r$ so that $\omega_{A\phi} \propto B_{\phi}/r\sqrt{\rho}$ is constant
(note that we have assumed constant density in the magnetosphere),
$[\ln(r^3\sigma^2)]'$ in Eq.~(\ref{eq:odem3}) reduces to
$[\ln(r^3\tomega^2)]'$. Therefore magnetic fields in
Eq.~(\ref{eq:odem3}) (which enters through term $\sigma^2$) completely
disappear and the equation is exactly the same as the one for
$\Omega=\mbox{const}$ and $B_{\phi}=0$. So the solutions for the
magnetosphere region in this case are also $\xi_r \propto r^{m-1}$,
$\delta h \propto r^{m}$ and the relation between these two
wavefunctions can be obtained by substituting them back into
Eqs.~(\ref{eq:odem1})-(\ref{eq:odem2}). Our calculations will focus
on this particular magnetosphere setup.

\subsection{The disc}

In the disc region ($r > r_{\rm in}$), we assume the flow is
barotropic so that $\delta P=c_s^2\delta \rho$ with
$c_s=\sqrt{dP/d\rho}$ being the sound speed, which will be
parametrized by a constant 
${\hat c_s}\equiv c_s/(r\Omega)$ in our
computation. For concreteness, we will also assume a power-law disc
density profile $\rho \propto r^{-p}$ with $p$ being a constant.
For simplicity, we take the disc toroidal magnetic field to be
uniformly distributed, i.e., $B_{\phi}={\rm const}$. From
Eq.~(\ref{eq:fb}), we obtain the disc rotation profile as 
\be
\Omega(r)\simeq\frac{\Omega_{\rm K}(r)}{\sqrt{1+p{\hat c_s}^2}}.
\ee
In the above expression, we have ignored the effect of magnetic field as the disc field is supposed to be fairly weak in our model (see Fig.~\ref{fig:interface}). The perturbation equations in component form can also be presented in
the same form as Eqs.~(\ref{eq:odem1}) and (\ref{eq:odem2}) with the
detailed expressions of $A_{11},A_{12},A_{21}$ and $A_{22}$ 
given in Fu \& Lai (2011a). For a non-magnetic disc, these equations
reduce to
\be 
\frac{d\xi_r}{dr}=\left[-\frac{2m\Omega}{r\tomega}-\frac{d\ln(r\rho)}{dr}\right]\xi_r+
\left(\frac{m^2}{r^2\tomega^2}-\frac{1}{c_s^2}\right)\delta h,
\label{eq:oded1}
\ee
\be
\frac{d\delta h}{dr}=(\tomega^2-\kappa^2)\xi_r+\frac{2m\Omega}{r\tomega}\delta h.
\label{eq:oded2}
\ee

\subsection{The interface}

In the equilibrium state, pressure balance at the interface reads
\be
P_{m}+P_{bz}+P_{b\phi}=P_d,
\ee
where $P_m$, $P_{bz}$ and $P_{b\phi}$ are the magnetosphere gas
pressure, magnetic pressure of $B_z$ and magnetic pressure of
$B_{\phi}$ just inside the interface, respectively, while $P_d$ is the
disc total pressure just outside the interface. This
equality imposes an upper limit on the strength of the magnetosphere
toroidal B field,
$P_{b\phi}/P_d<1$.
Defining $b=(\omega_{A\phi})|_{r_{\rm in }}/\Omega_d$ and
$\mu=(\rho_d-\rho_m)/(\rho_d+\rho_m)$, where $\Omega_d$ is the disc
rotation rate at the interface, $\rho_m$ and $\rho_d$ are the fluid
densities of the magnetosphere and the disc at the interface,
respectively, then the above inequality becomes
\be
\frac{P_{b\phi}}{P_d}\simeq\frac{(1-\mu)b^2}{(1+\mu)2{\hat c_s}^2}
\label{eq:pratio}
\ee
Note that in the magnetosphere, magnetic pressure dominates over gas
pressure ($P_{bz}+P_{b\phi} \gg P_m$). So $P_{b\phi}/P_d \simeq
P_{b\phi}/(P_{b\phi}+P_{bz})$ approximately characterizes the relative
strength of the toroidal field compared to the vertical field.

In the perturbed state, we demand that both the radial Lagrangian
displacement $\xi_r$ and the Lagrangian perturbation of total pressure
$\Delta \Pi=\delta \Pi+\xi_r\Pi'$ be continuous across the
interface. The latter of these requirements yields
\be
\rho_m \left[\delta h+\xi_r r (\Omega^2-\Omega_{\rm K}^2-\omega_{A\phi}^2)\right]_{r_{\rm in-}}=
\rho_d \left[\delta h-\xi_r r p{\hat c_s}^2\Omega^2\right]_{r_{\rm in+}}.
\label{eq:bc1}
\ee
In the magnetosphere, we have the analytical solutions for $\delta h$
($\propto r^m$) and $\xi_r$ ($\propto r^{m-1}$) (see Sec.~2.1),
which are related via
\be
\delta h=\frac{r}{m}\left[\tomega^2+2\tomega\Omega+m(2-m)\omega_{A\phi}^2\right]\xi_r.
\ee
We substitute this relation into the left hand side of
Eq.~(\ref{eq:bc1}) and consolidate it with another requirement
(continuity of $\xi_r$) to obtain the matching condition across the
interface
\be
\frac{(\omega-m\Omega_m)[\omega-(m-2)\Omega_m]}{m}+\Omega_m^2
=\frac{1+\mu}{(1-\mu)r_{\rm in}}\frac{\delta h}{\xi_r}+\left[1-\frac{2\mu p 
{\hat c_s}^2}{1-\mu}+(m-1)b^2\right]\Omega_d^2,
\label{eq:match}
\ee
where $\delta h$ and $\xi_r$ are disc solutions at the interface,
$\Omega_m$ and $\Omega_d$ are the rotation rates of the magnetosphere
and the disc at the interface, respectively.
In the case of $B_\phi=0$ in the magnetosphere,
this matching condition reduces to Eq.~(22) in Tsang \& Lai (2009b).
Note that Eq.~(\ref{eq:match}) has no explicit dependence on the disc magnetic  
field. This has to do with the fact that we chose an uniformly distributed disc B field. 

\section{Interface modes: Newtonian potential}

Li \& Narayan (2004) and Tsang \& Lai (2009b), who considered
incompressible and compressible discs respectively, have shown that
interface modes of magnetosphere-disc boundary can
be subject to Rayleigh-Taylor and/or Kevin-Helmholtz 
instabilities, depending on the density contrast and velocity shear
across the interface. In order to investigate the effects of
magnetosphere toroidal magnetic fields on these instabilities, in this
section we assume the disc region to be magnetic-field free and employ
the standard shooting method (Press et al. 1992) to solve
Eqs.~(\ref{eq:oded1})-(\ref{eq:oded2}) using the matching condition
Eq.~(\ref{eq:match}) as the inner boundary condition at $r_{\rm in}$
and outgoing wave condition (Tsang \& Lai 2009b) at the outer boundary
of the disc ($r_{\rm out}$). The complex mode frequency 
$\omega$ can then be determined as an eigenvalue of the system.
Note that in general we also need to apply shooting method in 
the magnetosphere region (see Appendix A), but for the specific magnetic 
field profile (and uniform rotation) of the magnetosphere 
we chose in this study, the perturbation equations in the magnetosphere 
have simple analytic solutions (see $\S2.2$) and 
the effect of magnetosphere on the interface modes 
is embodied by the interface matching condition 
(i.e, inner boundary condition for disc perturbation equations). 

In our computation, we use $p=1.5$, ${\hat c_s}=0.15$, 
and put the outer boundary at $r_{\rm out}=2.85r_{\rm in}$.
We vary the strength of magnetosphere toroidal magnetic field $b$ 
(therefore $P_{b\phi}/P_{d}$; Eq.~[\ref{eq:pratio}]) for different sets of
($\Omega_{m}/\Omega_{d}$, $\rho_{m}/\rho_{d}$) to see
how the eigenvalue $\omega$ will be modified. 
Throughout this section, we use Newtonian potential
$\Phi =-GM/r$, so that the Keplerian frequency is
$\Omega_{\rm K}=(GM/r^3)^{1/2}$, and the epicyclic frequency $\kappa$ equals 
$\Omega$.

\begin{figure}
\begin{center}
$
\begin{array}{cc}
\includegraphics[width=0.45\textwidth]{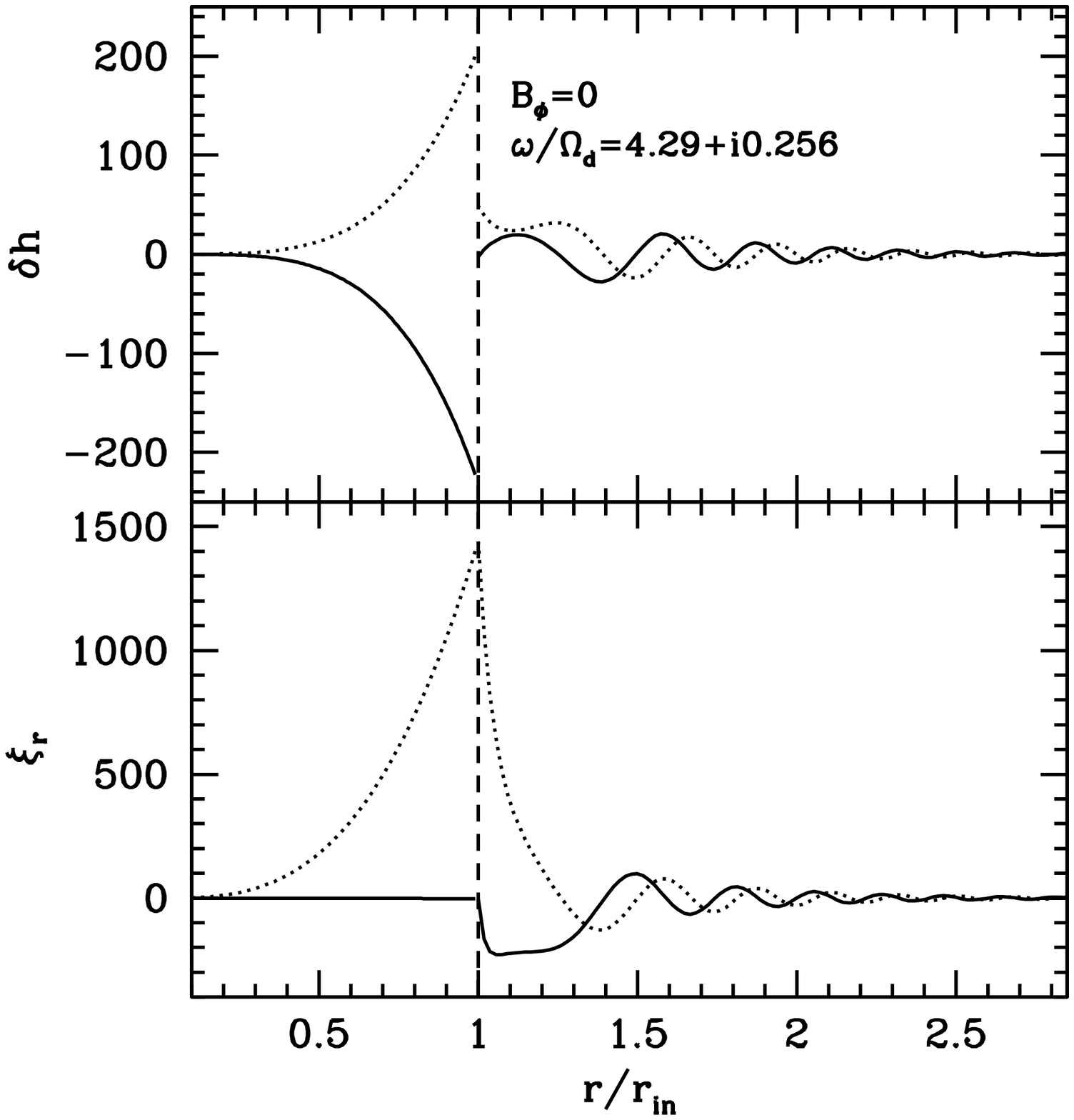} &
\includegraphics[width=0.45\textwidth]{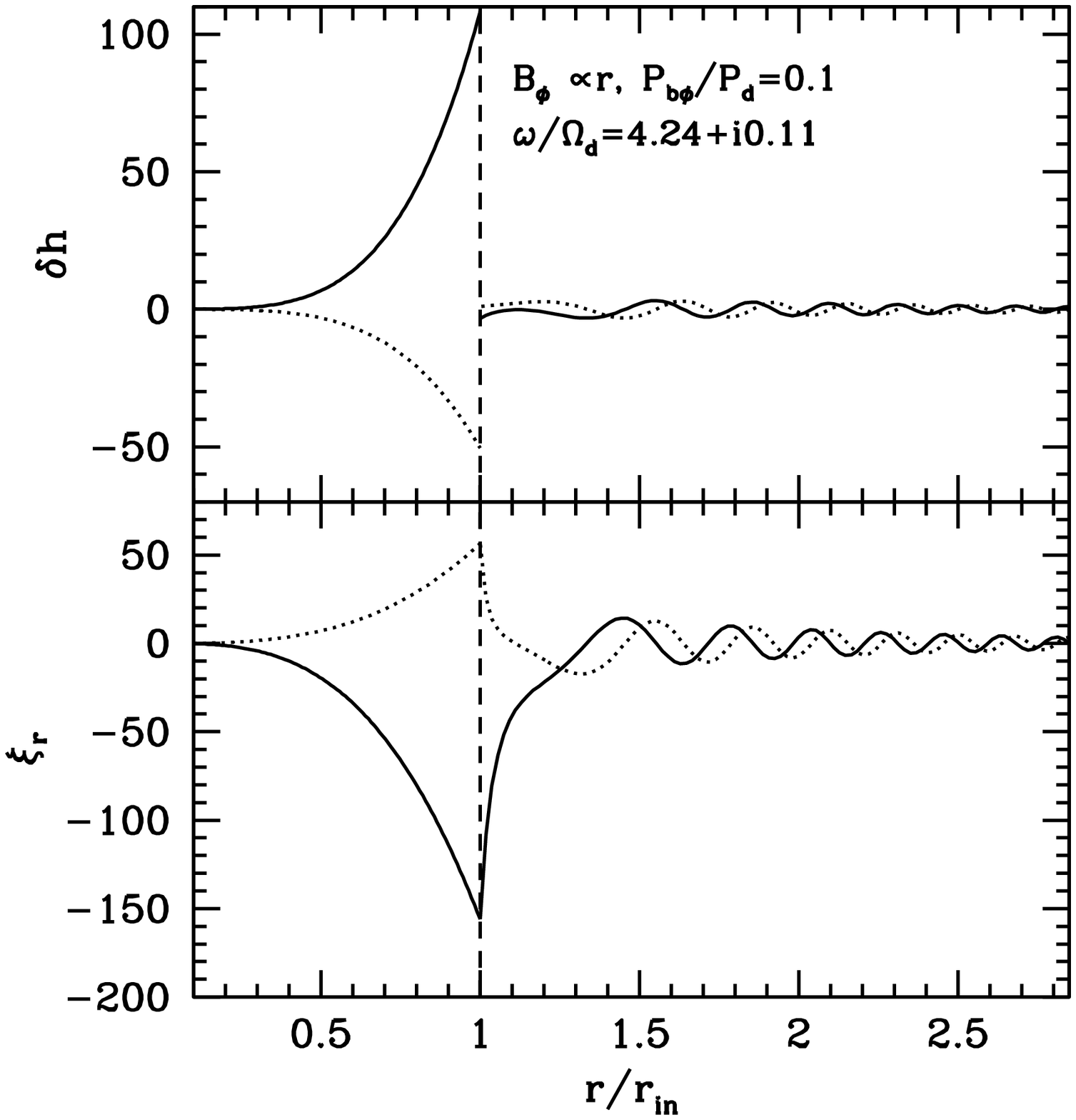}
\end{array}
$
\caption{The wavefunctions for the unstable interface mode with $m=4$,
  $\rho_{m}/\rho_{d}=1/99$, $\Omega_{m}/\Omega_{d}=1$ in a
  magnetosphere-disc system without (left) and with (right) magnetosphere   toroidal
  magnetic field, respectively. The vertical long-dashed lines mark
  the position of the interface which separates magnetosphere region
  and disc region. The solid and dotted lines represent the real and
  imaginary parts of the wavefunctions, respectively. }
\label{fig:wave}
\end{center}
\end{figure}

\begin{figure}
\begin{center}
$
\begin{array}{cc}
\includegraphics[width=0.45\textwidth]{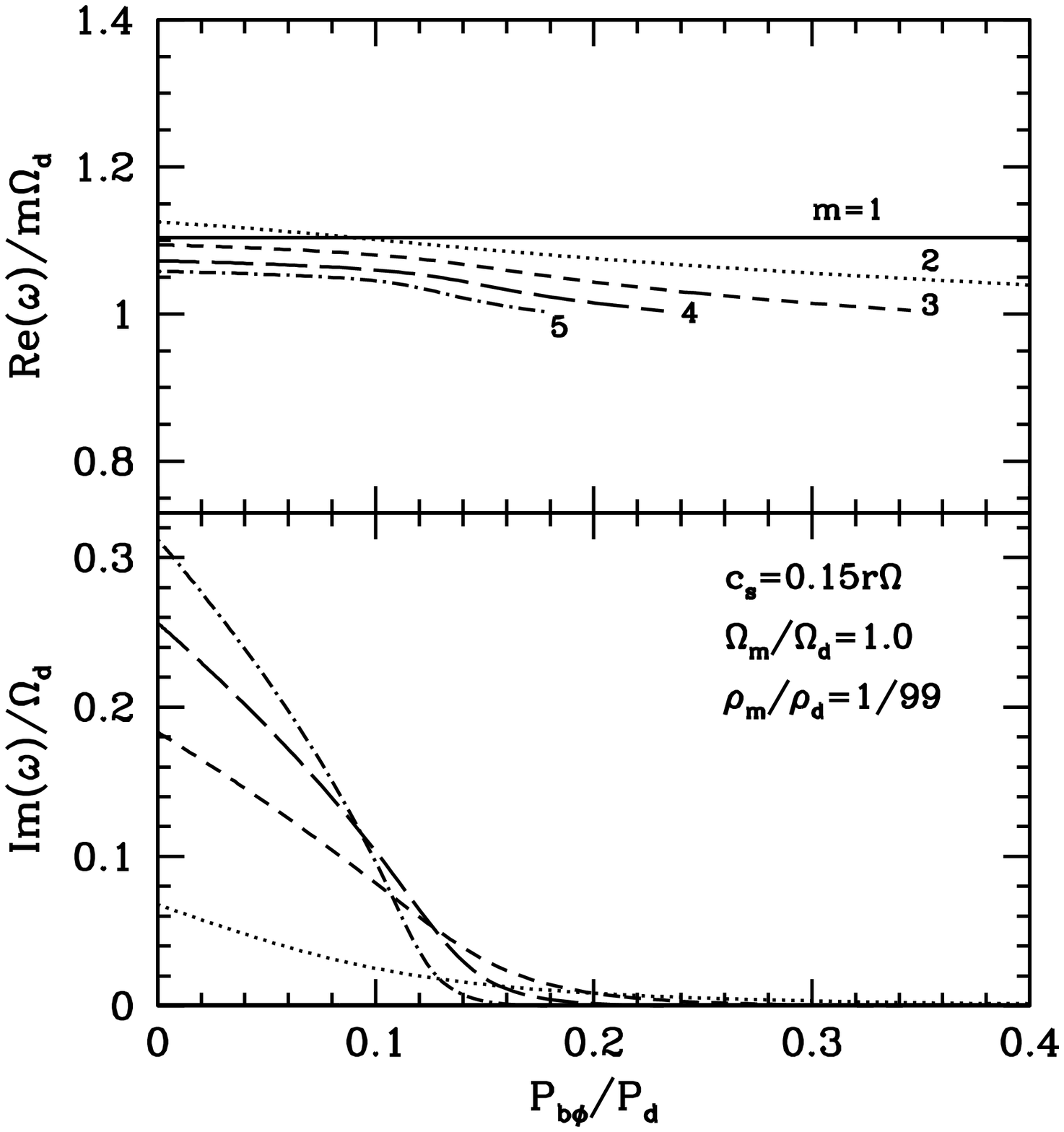} &
\includegraphics[width=0.45\textwidth]{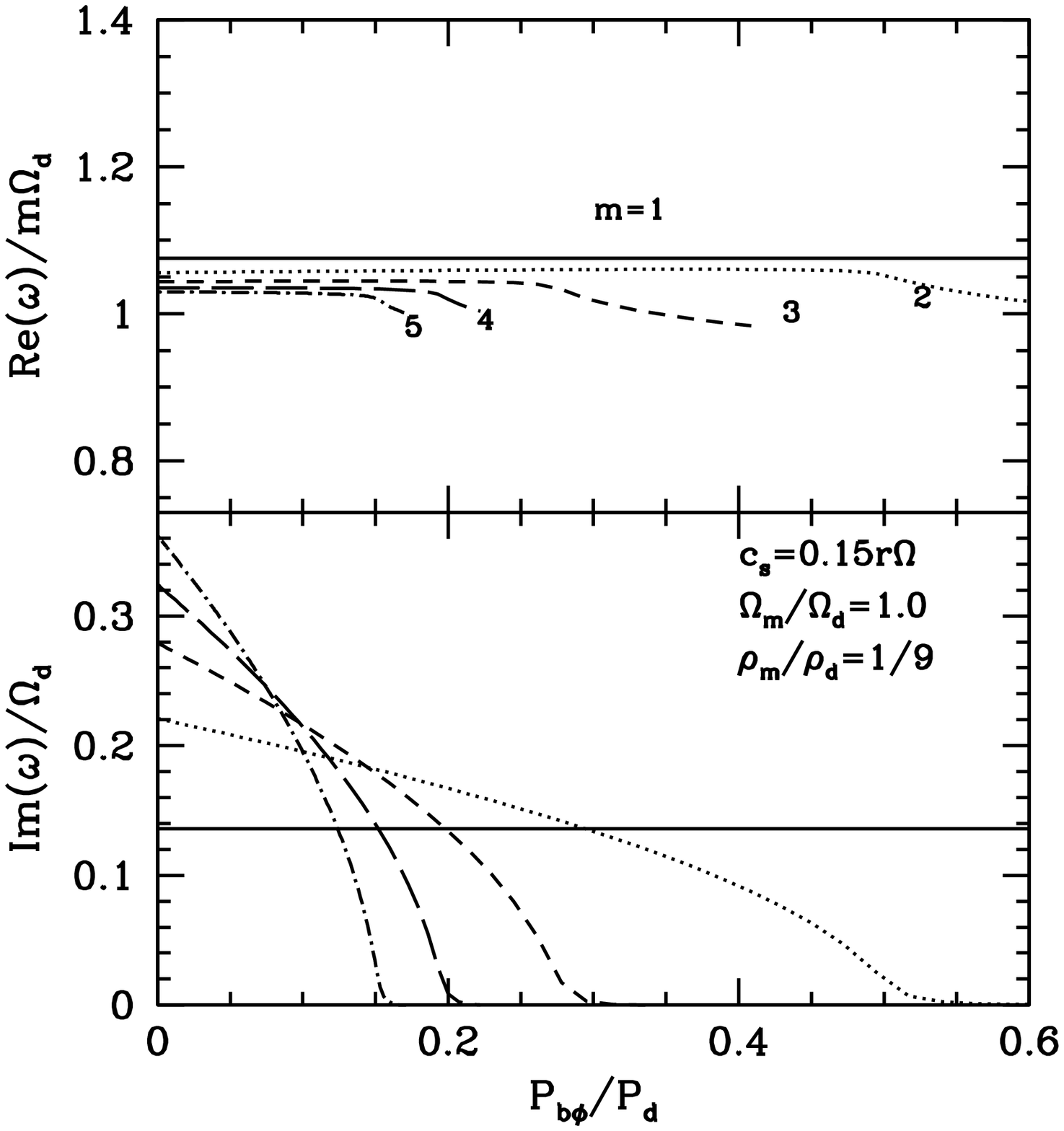} \\
\includegraphics[width=0.45\textwidth]{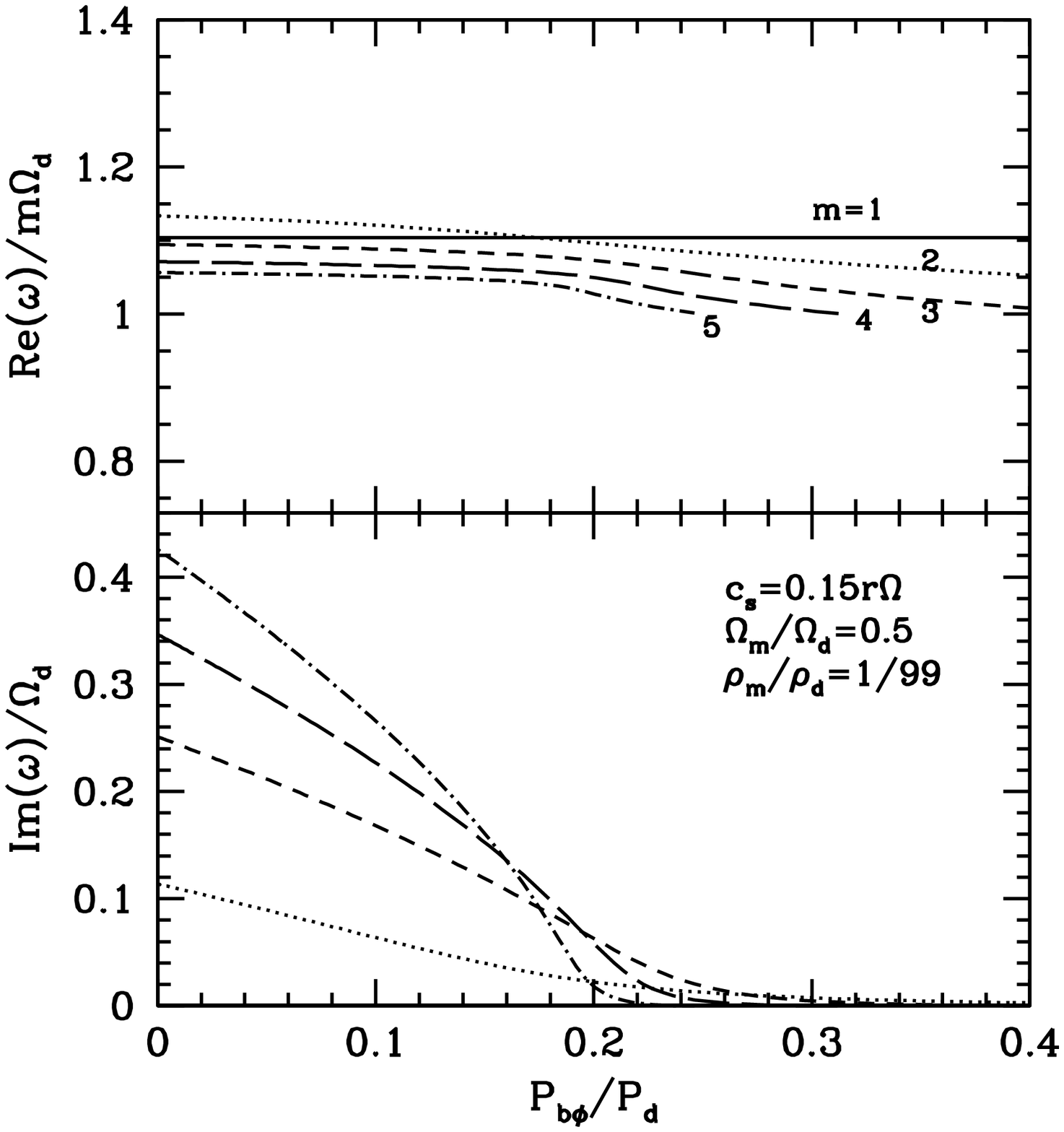} &
\includegraphics[width=0.45\textwidth]{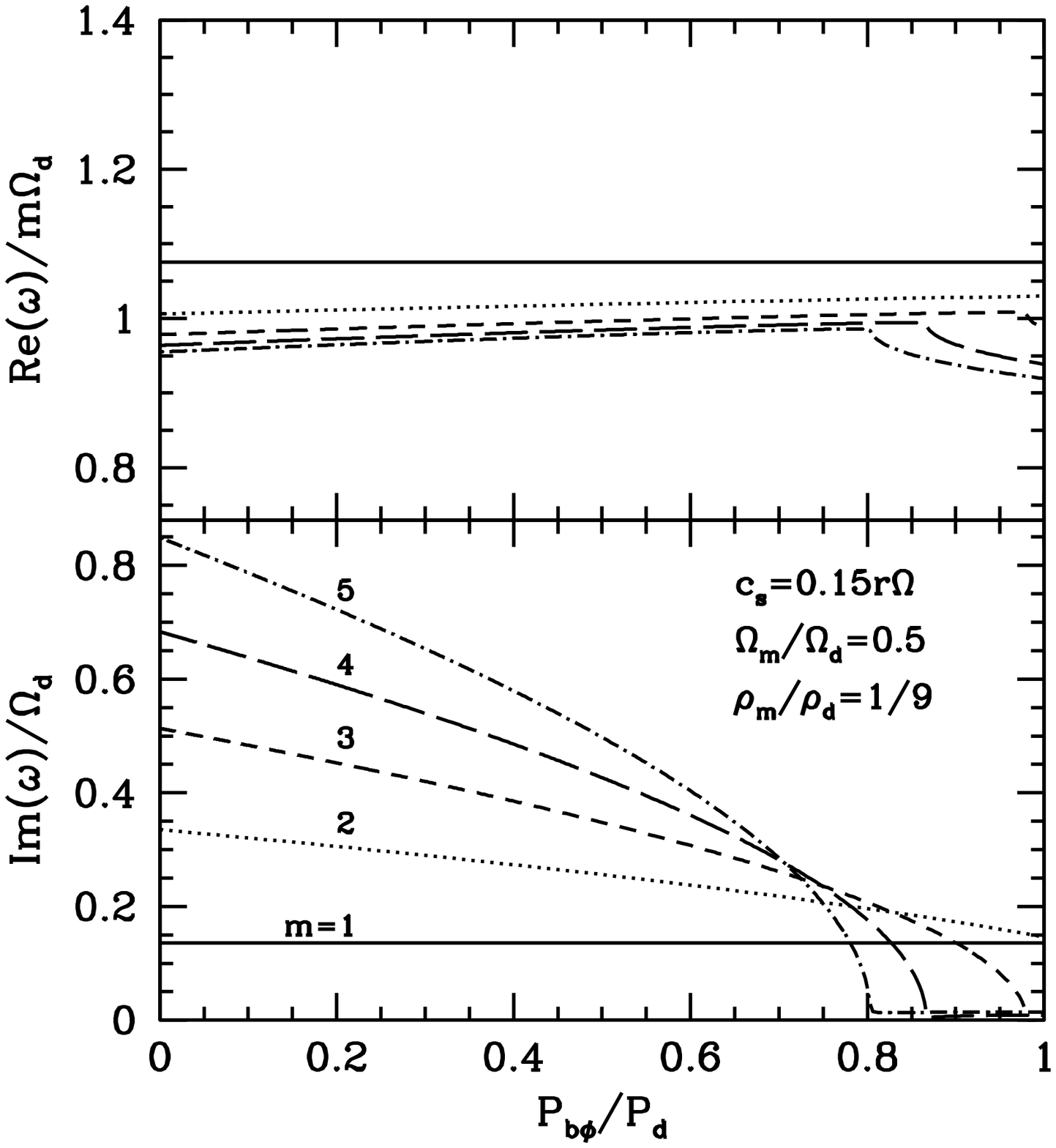}
\end{array}
$
\caption{Real and imaginary parts of the wave frequencies (in units of
  $\Omega_d$, the disc rotation rate at the interface) for unstable
  interface modes as a function of magnetosphere toroidal magnetic field
  strength. The top and bottom panels are for the cases
without and with
  velocity shear at the interface, respectively, while the left and
  right panels depict models with the magnetosphere to disc density
  ratio being $1/99$ and $1/9$, respectively. Different line types are
  associated with different values of $m$, with the solid lines for $m=1$,
  dotted lines for $m=2$, short-dashed lines for $m=3$, long-dashed
  lines for $m=4$, and dot-dashed lines for $m=5$. }
\label{fig:omega}
\end{center}
\end{figure}

\subsection{Numerical solutions}

Fig.~\ref{fig:wave} shows two example wavefunctions for the unstable
interface mode. The left panel depicts the case with no toroidal
magnetic field in the magnetosphere (but keep in mind that there is
always finite vertical B field) while in the right panel the
magnetosphere has a non-zero $B_{\phi}$. Since there is no velocity
shear across the boundary 
in both cases ($\Omega_m/\Omega_d=1$), the instability is
of Rayleigh-Taylor type. We see that 
the addition of toroidal field has a small effect on the eigenfunction.
However, as the numbers in the figure indicate, the growth rate of 
the unstable mode is reduced by more than 50\% (from 0.256$\Omega_d$ 
to 0.11$\Omega_d$) even though the the
toroidal magnetic field pressure is only 10\% of the disc gas pressure
(both evaluated at the interface), i.e. even though the toroidal field is
much weaker than vertical field in the magnetosphere.

The suppressing effect of toroidal magnetic field on the instability
can be more easily seen in Fig.~\ref{fig:omega}, where we plot the
real and imaginary parts of the eigen-frequencies for the interface
mode as a function of $B_{\phi}$. We consider four different sets of
$(\Omega_m/\Omega_d,~\rho_m/\rho_d$) such that the panels in the same
row have the same velocity shear but different density contrast while
the panels in the same column have the same density contrast but
different velocity shear. The modes in the top two panels are subject
to Rayleigh-Taylor type instability as there are no velocity shear at
the interface. The bottom two panels, however, have non-zero velocity
shear, thus are subject to instabilities of both Rayleigh-Taylor and
Kelvin-Helmholtz types. In all cases, we observe that the inclusion of
toroidal magnetic fields can significantly diminish the growth rate of
the unstable modes, even completely kill the instability. The critical
$B_{\phi}$ for absolute shut-down of unstable modes depends on the
detailed interface parameters (sound speed in the disc, velocity
shear, density contrast, etc.). In the case with the most unstable
modes (bottom-right panel) that we have calculated, the required
$P_{b\phi}/P_{d}$ to fully turn off the mode growth is close to
one. This means that the toroidal magnetic field needs to dominate
over the vertical field in the magnetosphere (remember that
$P_{b\phi}/P_{d}\simeq P_{b\phi}/(P_{b\phi}+P_{bz})$). The mode
frequencies (real part of $\omega$), on the other hand, are barely
affected by $B_{\phi}$ (at least when the growth rate is not very
small). By comparing panels in the same column, we see that increasing
velocity shear at the interface leads to larger growth rates, which is
not surprising as more velocity shear means more free energy that the
system can tap on to drive the instability. Comparing panels in the
same row reveals an interesting features. We found that larger density
contrasts leads to smaller growth rates, which apparently 
contradicts the standard Rayleigh-Taylor instability result. In
particular, the $m=1$ mode is totally stable when $\rho_m/\rho_d=1/99$
while becomes unstable when $\rho_m/\rho_d=1/9$ and $B_{\phi}$ does
not affect the $m=1$ mode at all. Comparing different lines in each
individual panel shows that modes with larger $m$ generally have
larger growth rates, but they are also more easily stabilized by
the toroidal field.

\subsection{Discussion of results}

The zero effect of $B_{\phi}$ on the $m=1$ mode (whether it is stable or
unstable) can be easily understood by looking at the matching
condition Eq.~(\ref{eq:match}). This equation (i.e., inner boundary
condition for disc equations) is the only place that $B_{\phi}$ 
affects the eigenvalue problem, and the magnetic effect comes in via
the last term on the right hand side of the equation
[$(m-1)b^2\Omega_d^2$]. When $m=1$, this term vanishes so that the
equation is the same as in the case with zero $B_{\phi}$. 

To explain
the other features observed in Fig.~\ref{fig:omega}, we 
carry out a local analysis of the effects of magnetic fields on 
Rayleigh-Taylor/Kelvin-Helmholtz instability in a plane-parallel flow 
(see Appendix B). From the last term in the final expression of wave frequency
(Eq.~[\ref{eq:solution}]), we see that magnetic tension
($\bb{k}\cdot\bb{B}$) provides a suppressing force against
Rayleigh-Taylor/Kelvin-Helmholtz instability of a two-layered fluid
system. This is because extra work needs to be done in order to
increase the boundary layer deformation (i.e., growing perturbation).
Our cylindrical magnetosphere-disc model resembles the simple parallel
two-layered flow in that our wavenumber in azimuthal direction $m/r$
acts like $k$ in the $x$ direction and our $B_{\phi}$ also lies along the
direction of the wave vector. 
Thus, the same stabilizing mechanism also applies in our system, 
i.e., the tension force of the toroidal magnetic field suppresses 
the magnetosphere-disc interface instability. 

Qualitatively, the growth rate of interface modes in our system
is determined by combined effects of four factors: density contrast
($\rho_m/\rho_d$), velocity shear across the interface ($\Omega_m/\Omega_d$),
degree of differential rotation in the disc, and toroidal field of the magnetosphere. 
The first two tend to enhance instability, while the latter two 
tend to suppress instability. In different limiting cases, approximate
analytic expressions for the mode growth rate can be derived: 
Li \& Narayan (2004) considered a case where the disc is 
incompressible with constant density and came up with their Eq.~(45),
while Tsang \& Lai (2009b) studied a compressible disc
whose density is much larger than the magnetosphere density, and
summarized the results in their Eqs.~(26) and (28). 
Although for the generic conditions studied in this paper, 
an analytical expression cannot be rigorously derived, 
we can write the mode growth rate schematically as follows:
\be  
\omega_{\rm i} \sim \sqrt{\omega_{\rm RT}^2+\omega_{\rm KH}^2-\omega_{\rm vort}^2-\omega_{b}^2}
\label{eq:growth}
\ee
with 
\be 
\omega_{\rm RT}^2\simeq 2(1+\mu)m\Omega_{\rm eff,d}^2-2(1-\mu)m\Omega_{\rm eff,m}^2,
\label{eq:growth1}
\ee
\be 
\omega_{\rm KH}^2\simeq (1-\mu^2)m^2(\Omega_{d}-\Omega_{m})^2+2(1-\mu^2)m(\Omega_{d}-\Omega_{m})(\zeta_{d}-\zeta_{m}),
\label{eq:growth2}
\ee
\be  
\omega_{\rm vort}^2\simeq (\zeta_{d}-\zeta_{m})^2+2\mu(\zeta_{d}^2-\zeta_{m}^2)+\mu^2(\zeta_{d}+\zeta_{m})^2,
\label{eq:growth3}
\ee
\be
\omega_{b}^2\simeq \frac{m^2B_{\phi}^2}{4\pi r^2(\rho_d+\rho_m)} \simeq (1-\mu)m^2\omega_{A\phi}^2/2,
\label{eq:suppress}
\ee
where $r\Omega_{\rm eff}^2 =d\Phi/dr-r\Omega^2$ is the effective
gravitational acceleration, $\zeta=\kappa^2/(2\Omega)$ is the fluid
vorticity and as before the subscripts $d$ and $m$ denote disc and
magnetosphere, respectively. Several features can be noted in
Eq.~(\ref{eq:growth}):

(i) The first two terms under the square root are the Rayleigh-Taylor and
Kelvin-Helmholtz types of destabilizing factors while the last two
terms characterize the stabilizing effects due to flow vorticity 
and toroidal magnetic field in the magnetosphere, 
respectively. The suppressing effect of finite fluid vorticity 
has been discussed in both Li \& Narayan (2004) and Tsang \& Lai
(2009b). We see that the various terms have different dependences on
the density contrast $\mu$.  As $\mu$ increases ($\mu=0$ when
$\rho_m/\rho_d=1$, and $\mu=1$ when $\rho_m/\rho_d=0$), $\omega_{\rm
  RT}^2$ increases, making the system more Rayleigh-Taylor unstable;
at the same time, $\omega_{\rm KH}^2$ decreases, making the system
less Kelvin-Helmholtz unstable.  In addition, $\omega_{\rm vort}^2$
also becomes larger, leading to stronger suppressing effect. This
explains why in Fig.~\ref{fig:omega}, the mode with
$\rho_m/\rho_d=1/99$ is less unstable than the one with
$\rho_m/\rho_d=1/9$.

(ii) Since $\Omega_{\rm eff,d}^2=(1/r)(d\Phi/dr)-\Omega^2=-(c_s^2/\rho)
(d\rho/dr)$, when $c_s$ is too small (i.e., small effective gravity in 
the disc), the destabilizing terms would not be able to compete with the
stabilizing terms, resulting in a stable system. Thus, 
a sufficiently high disc sound speed is needed to attain the interface 
instability (see Tsang \& Lai 2009b).

(iii) When there is no $B_{\phi}$, we see that $\omega_{\rm RT}^2$ and
$\omega_{\rm KH}^2$ both depend on $m$ while $\omega_{\rm vort}^2$ 
does not. Thus, modes with higher $m$ tend to be more
unstable, as shown in Fig.~\ref{fig:omega} (the $P_{b\phi}/P_d=0$ case;
see also Li \& Narayan 2004 and Tsang \& Lai 2009b). 
The $m=1$ mode
would be even less unstable if the density
contrast is too big [see (i) above]. This explains why
in Fig.~\ref{fig:omega} the $m=1$ mode with $\rho_m/\rho_d=1/99$ is
stable while the one with $\rho_m/\rho_d=1/9$ is unstable.

(iv) When $B_{\phi}\neq 0$, besides $\omega_{b}^2$, the magnetic field
strength also appears in $\omega_{\rm RT}^2$ where $\Omega_{\rm eff,m}^2$
contains a term that is proportional to $-B_{\phi}^2/4\pi\rho_m r^2$
(note the minus sign). This shows that 
the toroidal field in the magnetosphere plays two 
different roles in determining the stability of system. On the one hand, 
the magnetic tension resists perturbation growth, thus suppressing any
instability; on the other hand, the magnetic force 
increases the effective gravity (pointing towards the center) of the
background flow, thus promoting Rayleigh-Taylor type instability. The
former effect is proportional to $m$ while the latter to $m^2$. Hence,
in general the suppressing effect is more important.  This is
consistent with the fact that in Fig.~\ref{fig:omega} the growth rates
of modes with higher $m$ decrease more rapidly as $B_{\phi}$ increases.
As noted above, $B_{\phi}$ does not affect the $m=1$ mode because the
last term in Eq.~(\ref{eq:match}) vanishes for $m=1$. Now we have a
better understanding of what this means physically: For the $m=1$
mode, the aforementioned two opposing effects associated with
$B_{\phi}$ happen to cancel each other. This exact cancellation,
however, cannot be captured by the approximate expression
Eq.~(\ref{eq:growth}).  

Overall, we see that Eq.~(\ref{eq:growth}), though schematic, is quite
useful in explaining most of the numerical results of this section.
Note that various flow parameters, such as 
the density contrast $\mu$, azimuthal mode number $m$,
toroidal magnetic field $B_{\phi}$ and disc vorticity $\zeta_{d}$,
appear in more than one of the four terms in this equation. 
Thus, they are associated with both the stabilizing and
destabilizing effects. The only exception is perhaps
the velocity shear at the interface, which always facilitates
the interface instability. 

Finally, we note that the interface instability associated with
magnetosphere-disc boundary studied in this paper is qualitatively
different from those of Lubow \& Spruit (1995) and Spruit, Stehle \&
Papaloizou (1995), who carried out local analysis of a thin rotating
disc threaded by a large-scale poloidal field. 
Lovelace, Romanova \& Newman (2010) considered similar interface
instability as in our paper, but they focused on small-scale
modes and included only vertical field in the magnetosphere.

\section{Interface modes and Discoseismic modes: Pseudo-Newtonian potential}

In this section, we consider how the global oscillation modes in our
disc-magnetosphere system (Fig.~1) are modified by general
relativistic (GR) effects. In particular, GR changes the disc rotation
profile $\Omega_{\rm K}(r)$ and makes the radial epicyclic frequency
$\kappa(r)$ a non-monotonic function of $r$: As $r$ decreases,
$\kappa$ first increases, attains a maximum value and then falls
to zero at the Innermost Stable Circular Orbit (ISCO), $r_{\rm
  ISCO}=6GM/c^2$ (for a non-spinning compact object).  If the
inner disc boundary is close to $r_{\rm ISCO}$, then this
non-monotonic $\kappa$ profile can have two consequences: (1) It
significantly reduces disc vorticity ($\kappa^2/2\Omega$) near the
inner disc boundary, therefore helps the unstable interface modes grow
as there are less disc vorticity suppressing effects (see the discussion
in Section 3.2; see also Tsang \& Lai 2009b);  (2) It could lead to a
vortensity ($\kappa^2/2\Omega\rho$ a.k.a potential vorticity) profile
that has positive gradient in the inner disc region. As found in our
previous studies (Tsang \& Lai 2008; Lai \& Tsang 2009; Fu \& Lai
2011a), positive vortensity gradient renders the disc
inertial-acoustic modes (p-modes) unstable.  In Section 3, we chose a
density profile $\rho \propto r^{-3/2}$ so that the vortensity profile
(with the Newtonian potential) is completely flat, and thus we only 
found unstable (overstable) interface modes. In this section, with the 
GR effect included, both the interface modes and disc p-modes can be 
unstable and they co-exist in our disc-magnetosphere system.
The main goal of this section is then to study these two modes and 
the effects of the magnetosphere toroidal B field, disc toroidal B field 
and inner disc boundary location. 

We employ the pseudo-Newtonian potential (Paczynski \& Wiita 1980) 
to mimic the GR effect:
\be 
\Phi=-\frac{GM}{r-r_{\rm S}},
\ee
where $r_{\rm S}=2GM/c^2$ is the Schwarzschild radius. The
corresponding Keplerian rotation frequency and epicyclic frequency are
\ba
&&\Omega_{\rm K}=\sqrt{\frac{GM}{r}}\frac{1}{r-r_{\rm S}},\\
&&\kappa=\Omega\sqrt{\frac{r-3r_{\rm S}}{r-r_{\rm S}}}.
\label{eq:grkappa}
\ea
Except for Section 4.3, we will take 
the magnetosphere-disc interface to be located at the ISCO,
$r_{\rm in}=r_{\rm ISCO}=3r_{\rm S}$.

\subsection{Non-magnetic discs}

We first consider the case where the disc outside the magnetosphere is
non-magnetic. We solve the same equations and use the same
interface boundary condition as in Section 3 except that we replace the
Newtonian potential with Pseudo-Newtonian potential.

\begin{figure}
\begin{center}
$
\begin{array}{cc}
\includegraphics[width=0.47\textwidth]{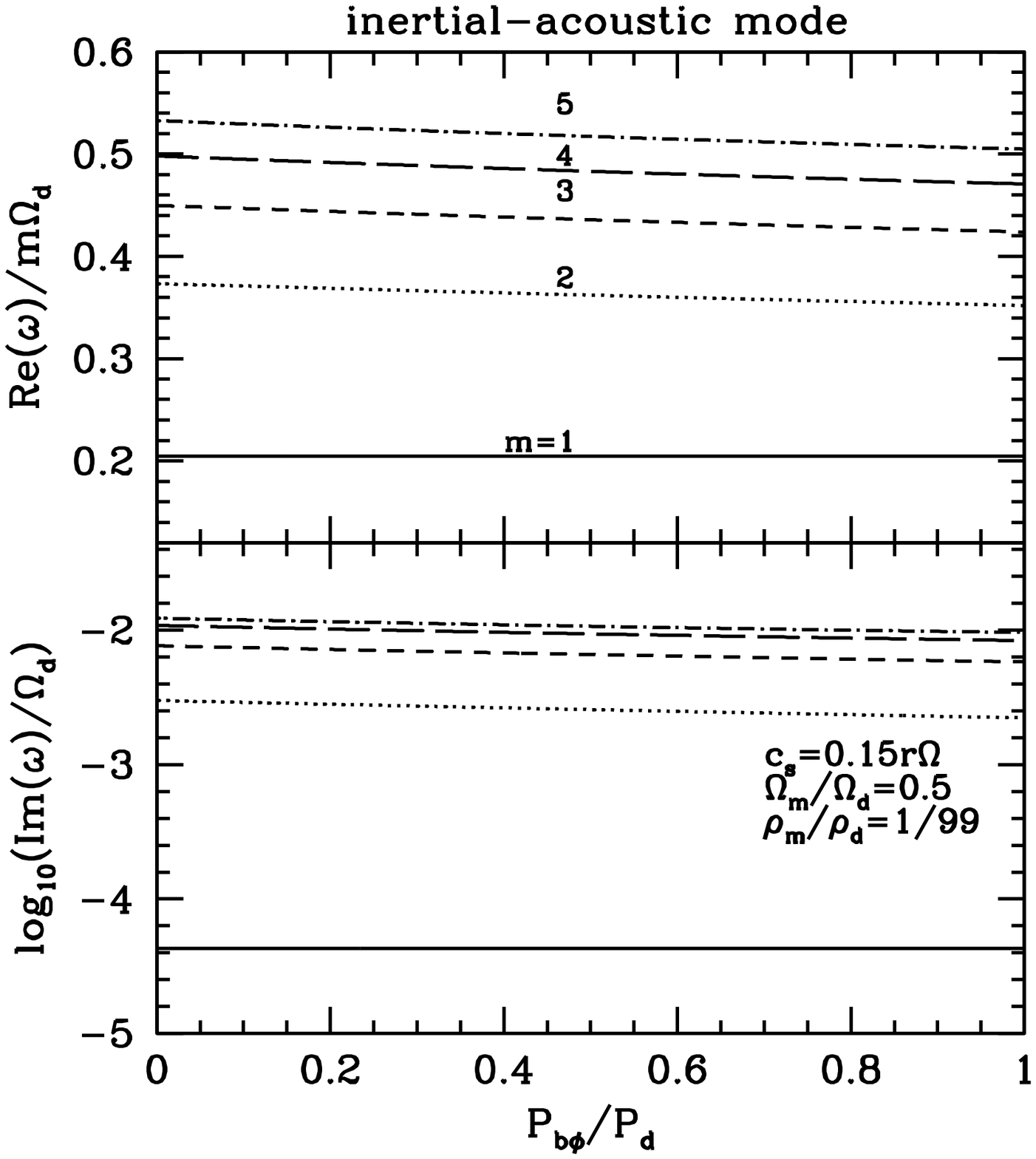} &
\includegraphics[width=0.47\textwidth]{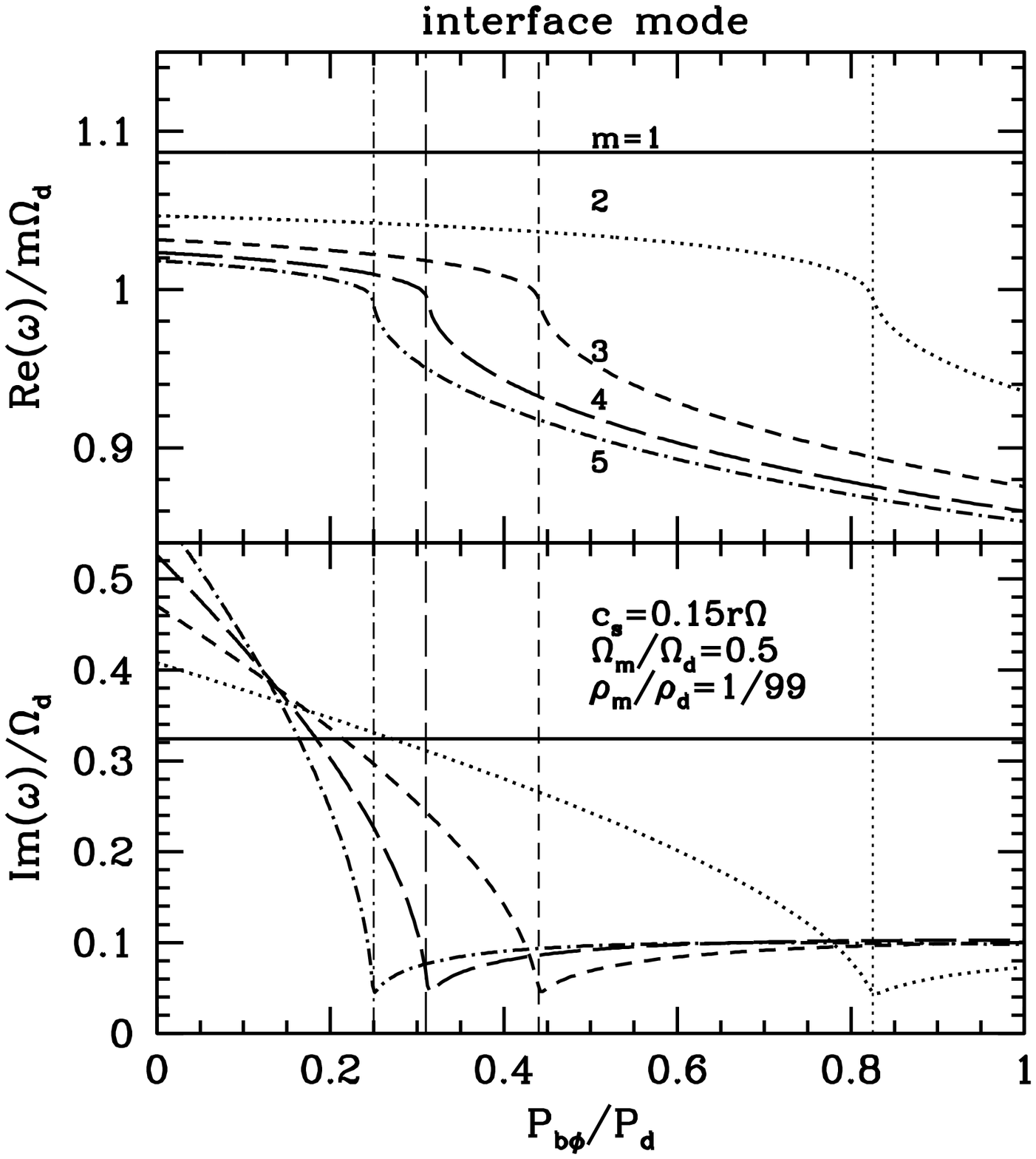}
\end{array}
$
\caption{Real and imaginary parts of the wave frequencies (in units of
  $\Omega_d$, the disc rotation rate at $r_{\rm in}$) for unstable
  inertial-acoustic disc modes (left panels) and unstable interface
  modes (right panels) as a function of the magnetosphere toroidal
  field strength. Different line types represent different azimuthal
  mode number $m$. The sound speed profile, density contrast and
  velocity shear across the interface are the same as in the
  bottom-left panel of Fig.~\ref{fig:omega}. The only difference from
  Fig.~\ref{fig:omega} is that GR effect (using pseudo-Newtonian
  potential) is included in the disc rotation profile, which renders
  both types of modes unstable. On the right panels, the vertical
  lines (for $m=2$ and 3) mark the point when corotation resonance 
  moves into the flow (see Fig.~\ref{fig:diagram}).}
\label{fig:pn_omega}
\end{center}
\end{figure}

\begin{figure}
\begin{center}
\includegraphics[width=0.6\textwidth]{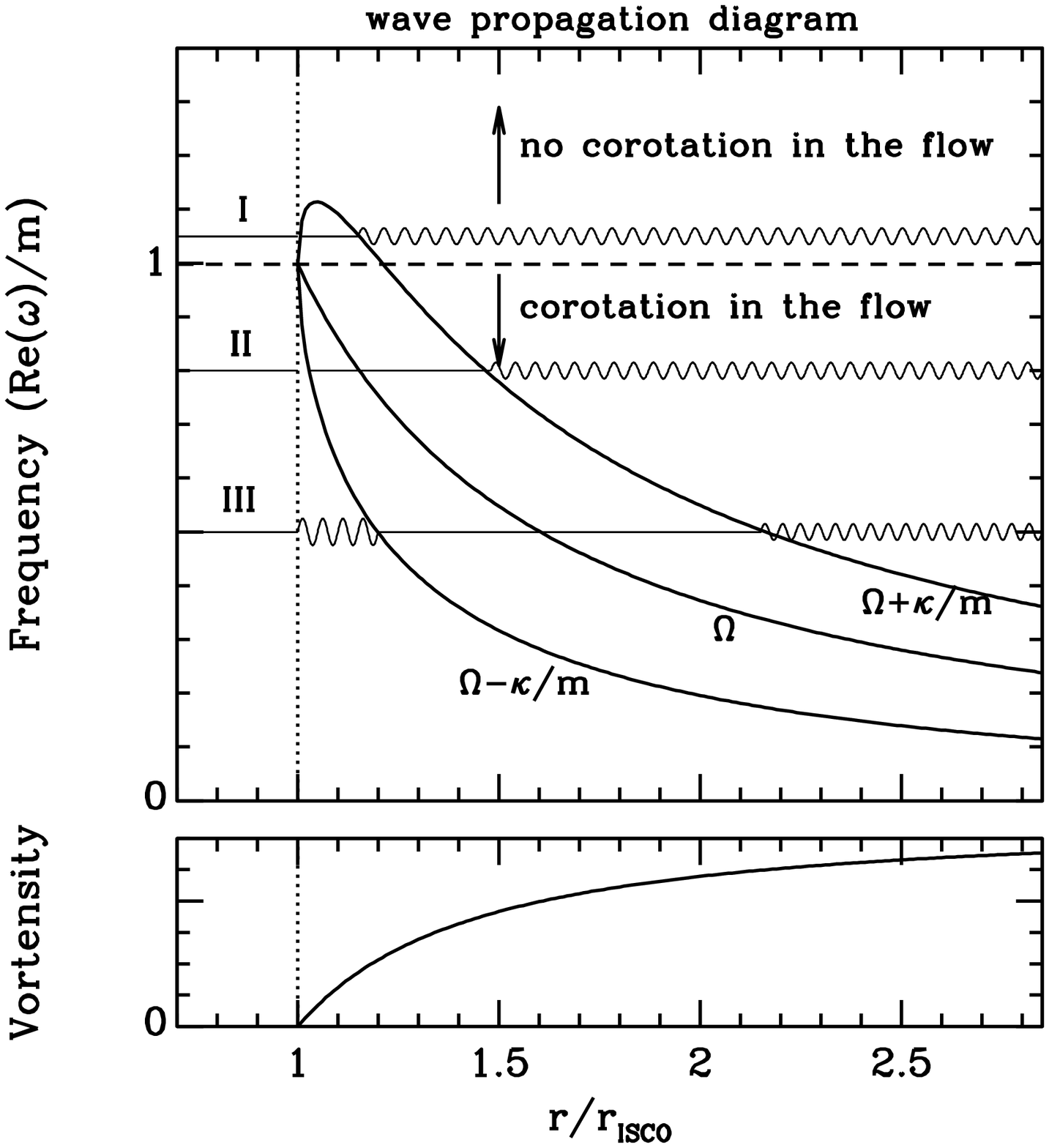}
\caption{Wave propagation diagram (upper panel) for both interface
  modes and disc inertial-acoustic modes accompanied by the disc
  vortensity profile (lower panel). In the upper panel, the three
  thick solid curves depict the disc rotation profile $\Omega$ and
  $\Omega\pm \kappa/m$, where $m$ is the azimuthal mode number and
  $\kappa$ is the radial epicyclic frequency. Note that since
  $\kappa(r_{\rm ISCO})=0$ these three curves join each other at the
  Innermost Stable Circular Orbit (which is also the disc inner
  boundary $r_{\rm in}$ in our setup of Sections 4.1 \& 4.2). The
  dashed horizontal line marks the point when the corotation radius
  [where ${\rm Re}(\omega)/m=\Omega$] happens to be exactly at the
  inner disc boundary. For modes with ${\rm Re}(\omega)/m$ {\it above}
  this line, there is no corotation in the flow; for any modes with
  ${\rm Re}(\omega)/m$ {\it below} this line, the corotation exists in
  the flow ($r_{\rm c} > r_{\rm in} $).  Three types of wave modes
  (labeled as I, II and III) are shown in this diagram where the
  horizontal solid lines indicate the evanescent zones and the wavy
  curves denote the wave propagation zones. Note that for both Type I
  and II waves, there is also a tiny wave zone (hardly visible in the
  figure) in the disc region just outside the ISCO. Type I and II
  waves are both interface modes while type III waves are disc
  inertial-acoustic modes. Note that the magnetosphere region ($r<
  r_{\rm ISCO}$) and the region between the inner and outer Lindblad
  resonances [where ${\rm Re}(\omega)/m=\Omega \pm \kappa/m$] are wave
  evanescent zones.}
\label{fig:diagram}
\end{center}
\end{figure}

\begin{figure}
\begin{center}
$
\begin{array}{ccc}
\includegraphics[width=0.33\textwidth]{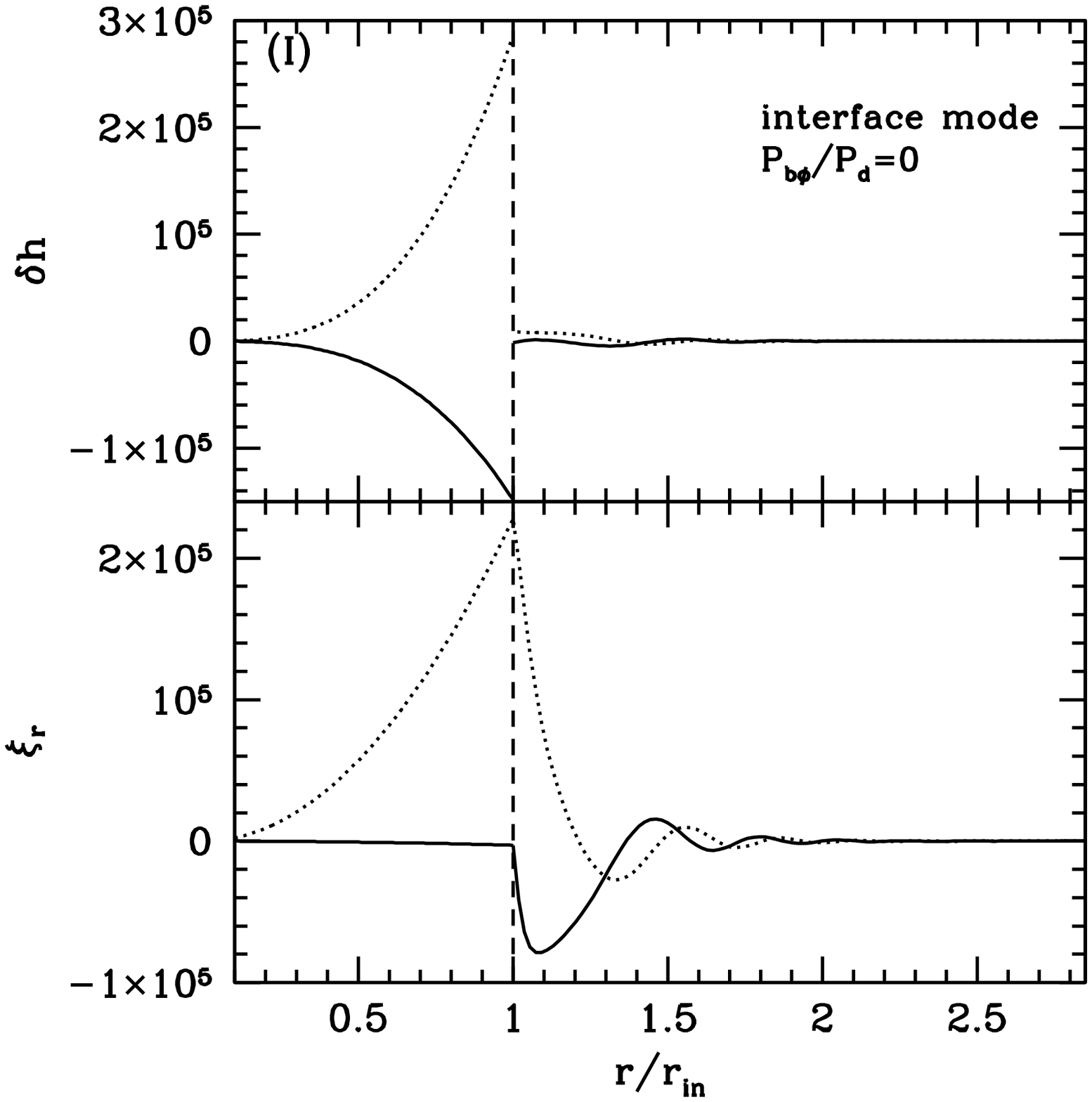} &
\includegraphics[width=0.33\textwidth]{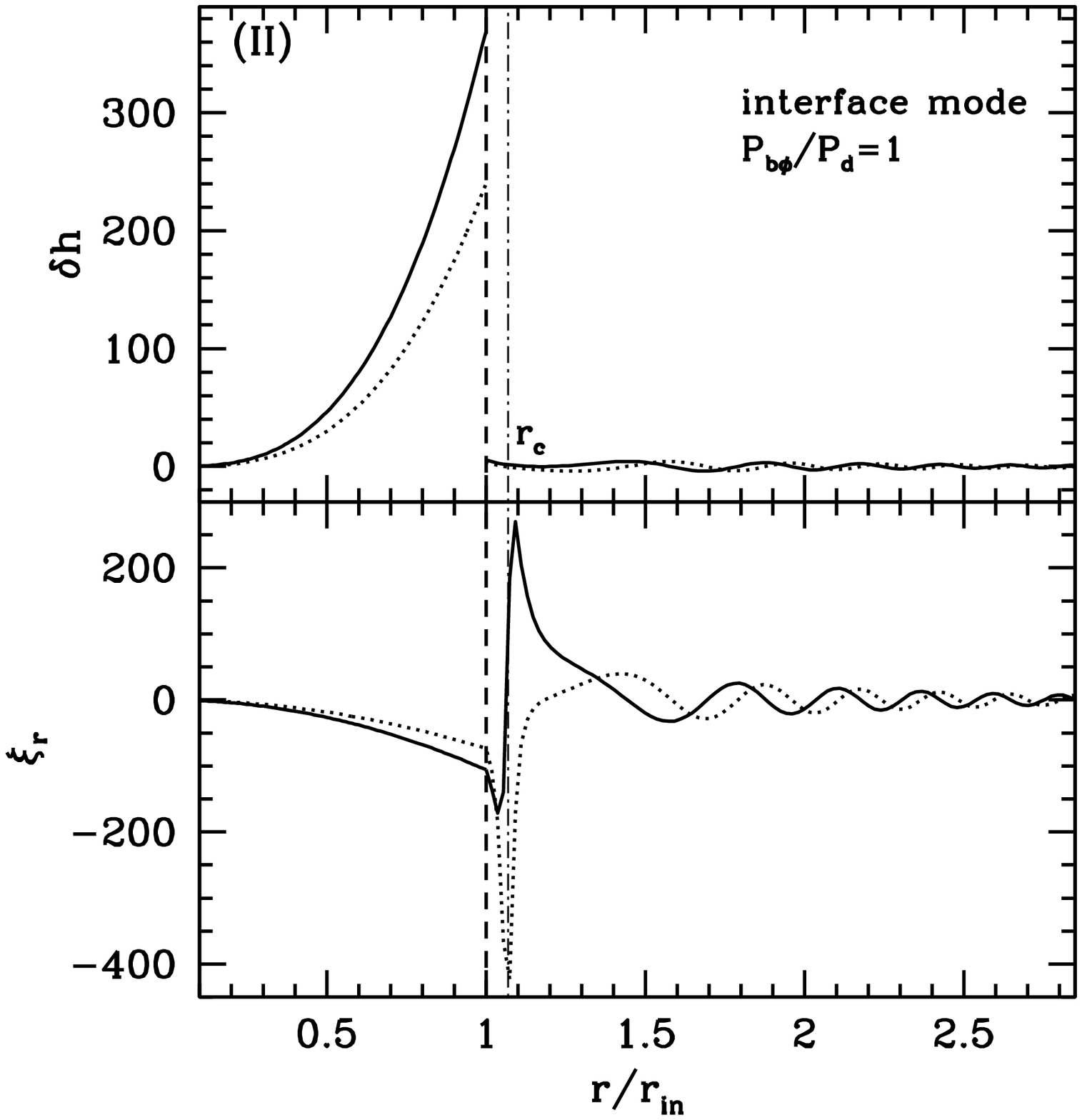} &
\includegraphics[width=0.33\textwidth]{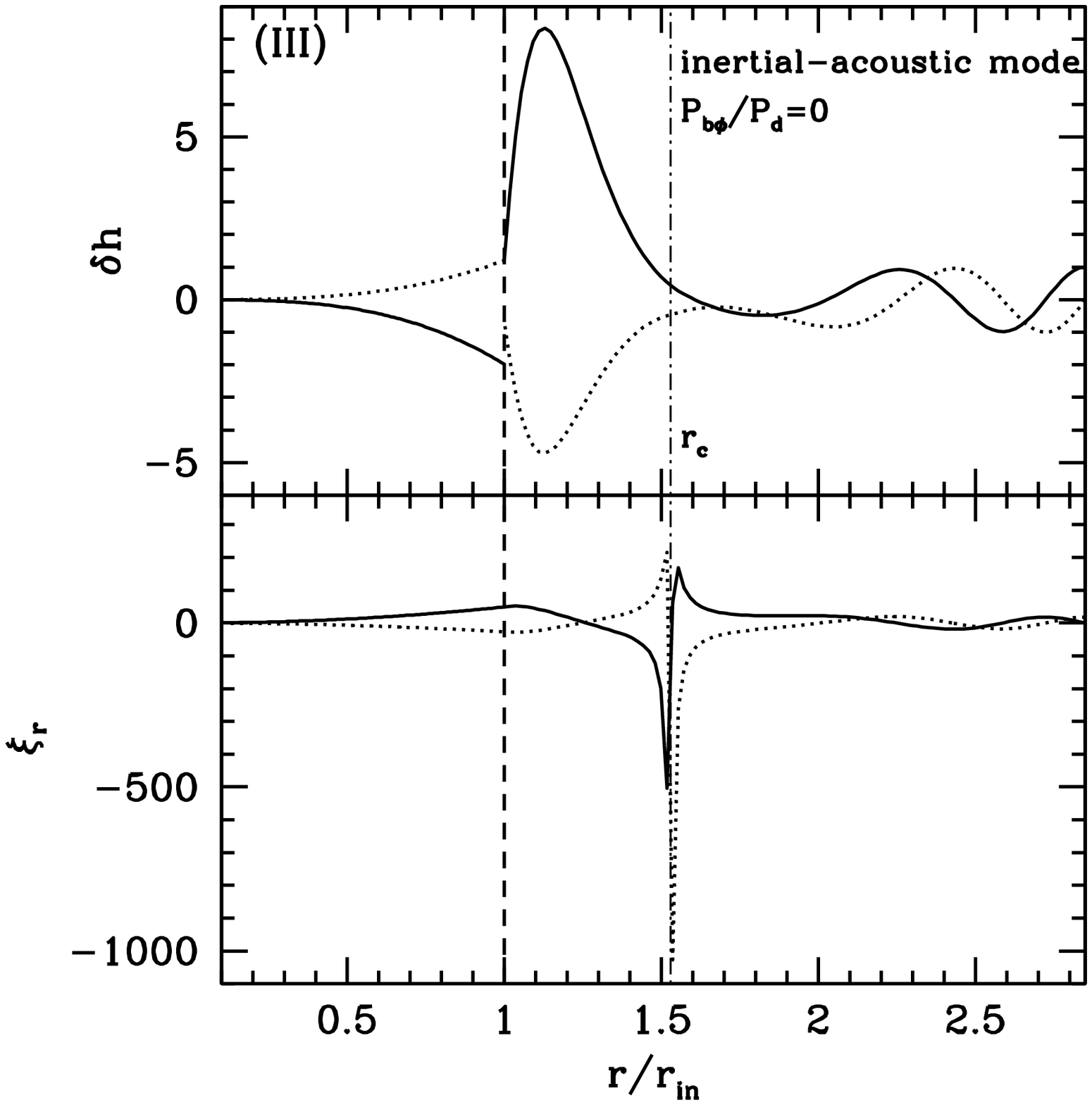}
\end{array}
$
\caption{Wave functions for the three types of wave modes
depicted in Fig.~\ref{fig:diagram}: the interface mode with 
no corotation resonance in the flow (Type I, left panels),
interface mode with corotation resonance in
the flow (Type II, middle panels) and disc inertial-acoustic mode 
(Type III, right panels). The azimuthal mode number is $m=3$ and all other
parameters are the same in Fig.~\ref{fig:pn_omega}. The solid
  and dotted lines show the real and imaginary parts, respectively. In
  the middle and right panels, the vertical dot-dashed lines represent
  the location of corotation resonance.}
\label{fig:pn_wave}
\end{center}
\end{figure}

Figure~\ref{fig:pn_omega} shows the complex eigenfrequencies of disc
inertial-acoustic modes and interface modes as a function of the
dimensionless magnetosphere toroidal field strength $P_{b\phi}/P_d$.
The other parameters are indicated in the figure.  As noted above,
because of the GR effect, the disc vortensity profile has a positive
slope near the ISCO (see Fig.~\ref{fig:diagram}).  The disc
inertial-acoustic mode (p-mode) of lowest radial order (for a given
$m$) has ${\rm Re}(\omega)<m\Omega_d$, and the corotation resonance
lies in the disc (see the right panels of Fig.~\ref{fig:pn_wave} for
an example).  Corotational wave absorption then makes these modes
unstable.  We see from Fig.~\ref{fig:pn_omega} that as long as there
is a magnetosphere to serve as an inner boundary condition for the
disc mode, the mode frequency and growth rate depend rather weakly on
$P_{b\phi}/P_d$.

The interface mode is also strongly influenced by the the GR effect.
Comparing the right panels of Fig.~\ref{fig:pn_omega} with the
bottom-left panel of Fig.~\ref{fig:omega}, we see that for
$P_{b\phi}=0$, the mode growth rate (for a given $m$) is larger when
the GR effect is included. Again, this is because in GR, the disc
vorticity is smaller near the ISCO, leading to less rotational
suppression of the RT instability (see Section 3.2). As
$P_{b\phi}/P_d$ increases, the interface mode growth rate first
decreases (just as in the Newtonian case) due to magnetic tension,
then starts to increase beyond certain critical value of
$P_{b\phi}/P_d$ (for a given $m$). This behaviour arises
from the effect of corotation resonance 
(see Fig.~\ref{fig:diagram}): For small 
$P_{b\phi}/P_d$, the interface mode has ${\rm Re}(\omega)>m\Omega_d$,
thus no corotation resonance exists in the flow; as
$P_{b\phi}/P_d$ becomes larger, the mode frequency drops below $m\Omega_d$,
and corotation resonance comes into play, which overwhelms the suppression
effect of the magnetic tension (see the left and middle
panels of Fig.~\ref{fig:pn_wave} for two examples of mode wavefunctions).
The growth rate of such interface mode (labelled as Type II
in Fig.~\ref{fig:diagram}) is much larger than the corresponding disc p-mode 
because the corotation resonance lies much closer to $r_{\rm in}$ 
for the interface mode than for the p-mode -- this gives rise
to much stronger wave absorption at the corotation resonance.

\subsection{Magnetic discs}

Now we consider the case in which the disc outside the magnetosphere
has a finite toroidal magnetic field.  For simplicity, we take the
limit that the density in the magnetosphere $\rho_m$ goes to zero. In
this case, the interface boundary condition Eq.~(\ref{eq:match})
reduces to $\Delta\Pi=0$, i.e., the Lagrangian perturbation of total
disc pressure equals zero at $r_{\rm in}$. With this boundary condition, 
the setup is the same as in Fu \& Lai (2011a) and we only need
to consider the dynamical equations for the disc.

\begin{figure}
\begin{center}
$
\begin{array}{cc}
\includegraphics[width=0.47\textwidth]{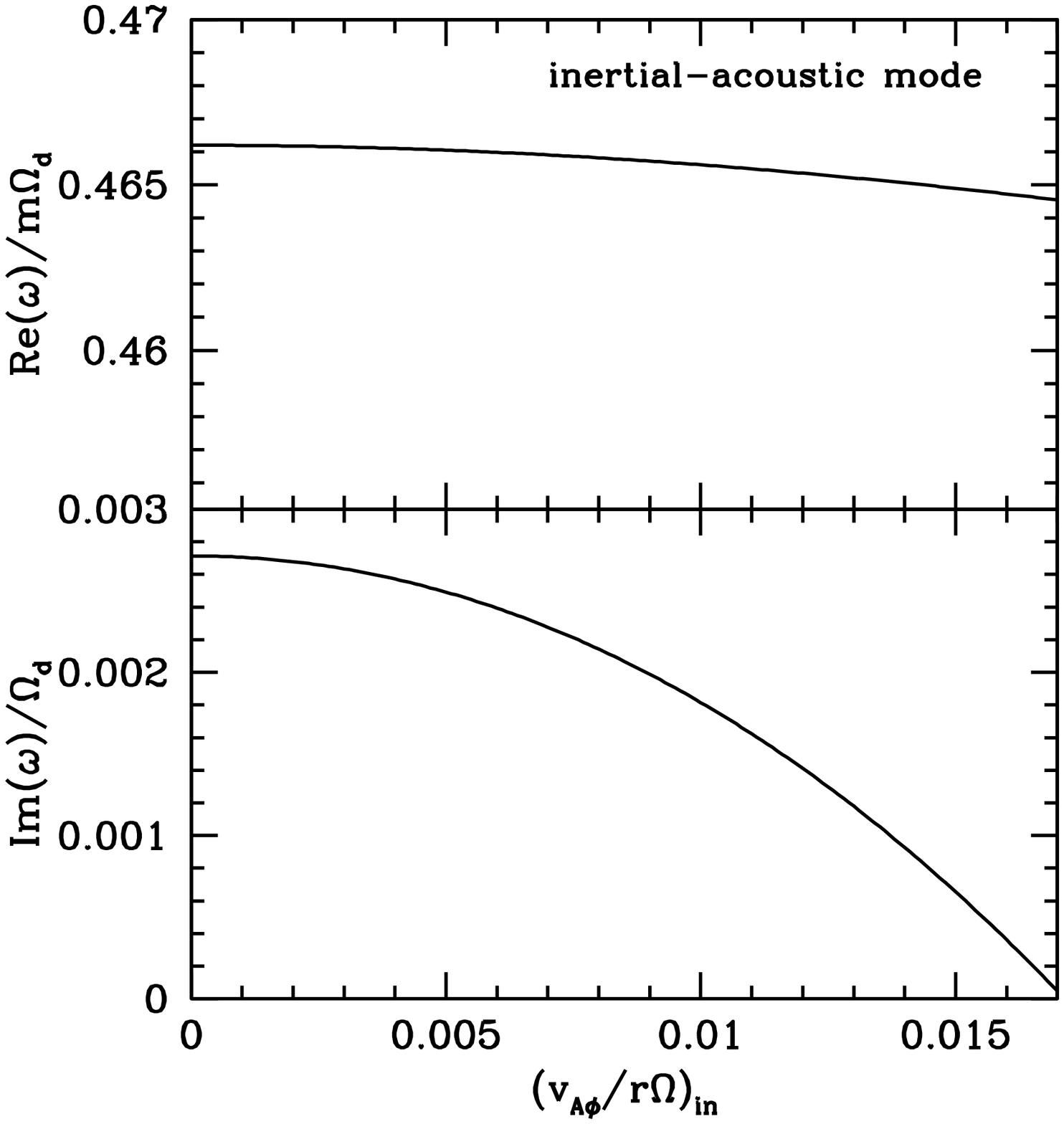} &
\includegraphics[width=0.47\textwidth]{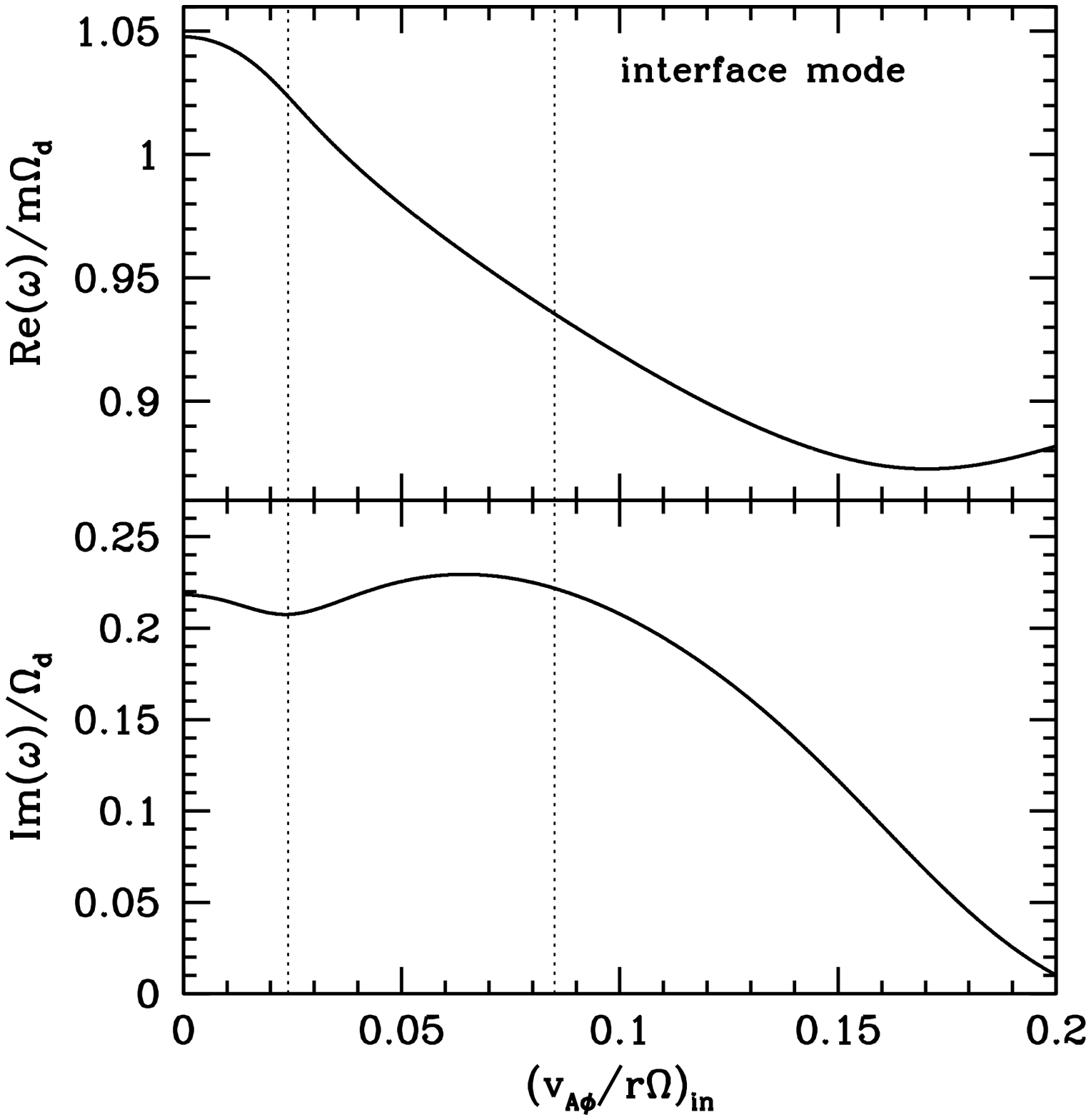}
\end{array}
$
\caption{Real and imaginary parts of the wave frequencies (in units of
  $\Omega_d$, the disc rotation rate at $r_{\rm in}$) for unstable
  inertial-acoustic disc modes (left panels) and unstable interface
  modes (right panels) as a function of the disc toroidal field
  strength. The $x$-axis specifies the ratio of the \Alfven velocity
  to disc rotation velocity at the inner boundary $r_{\rm in}$.  The
  magnetosphere density is set to zero. The disc parameters are
  $r_{\rm in}=r_{\rm ISCO}$, $m=2$, $c_{s}=0.1r\Omega$ and $\rho
  \propto r^{-1}$. For the interface mode (right panels), the two
  dotted vertical lines from left to right mark the entering of the
  outer magnetic resonance and inner magnetic
  resonance into the flow, respectively. }
\label{fig:fl_omega}
\end{center}
\end{figure}

Again, two types of oscillation modes exist in our system (see
Fig.~\ref{fig:fl_omega}): the disc inertial-acoustic modes and
interface modes. The effect of the disc magnetic field on the
inertial-acoustic modes has already been studied in detail by Fu \& Lai
(2011a). We see from the left panels of Fig.~\ref{fig:fl_omega}
that the mode frequency is only slightly modified by the disc $B_\phi$,
but the growth rate is reduced so that the mode become stable
even for modest (sub-thermal) disc toroidal fields.

The interface mode (the right panels of Fig.~\ref{fig:fl_omega})
was not considered by Fu \& Lai (2011a). We see that as
the disc magnetic field increases, the mode frequency 
varies modestly (about 10\%) for a wide range of $B_\phi$,
but the growth rate changes more significantly. 
The dependence of the growth rare as a function of
disc $B_\phi$ can be understood as follows: 
(i) For $B_\phi=0$, RT instability drives the mode growth;
(ii) As $B_\phi$ increases, the magnetic tension tends to suppress
the growth;
(iii) As $B_\phi$ increases, the real mode frequency
decreases. When $B_\phi$ exceeds some critical value, 
corotation resonance appears in the flow (disc), and 
wave absorption at corotation then enhances the growth rate. 
But as $B_\phi$ increases further, the corotation radius
lies at a larger distance from $r_{\rm in}$, thus 
the corotational effect becomes less important (because
of the large evanescent zone separating $r_{\rm in}$ and the corotation
radius) and the mode growth rate decreases again also because of magnetic
tension.

Concerning (iii) above, as noted in Fu \& Lai (2011a), in the 
presence of the disc toroidal magnetic field, the corotation resonance
(where $\tomega=0$) is split into the inner/outer magnetic resonances, where
\be
\tomega=\pm m\omega_{A\phi},
\ee
where $\omega_{A\phi}=v_{A\phi}/r=B_\phi/(r\sqrt{4\pi\rho})$ is the
toroidal \Alfven frequency of the disc. When these magnetic resonances
exist in the flow (disc), wave absorption comes into play in the mode
growth. Note that the signs of wave absorption at the two magnetic
resonances are different. As ${\rm Re}(\omega)$ decreases (with
increasing $B_\phi$), the outer magnetic resonance (where
$\tomega=m\omega_{A\phi}$) first enters the flow, causing the mode
growth rate to increase. When the inner magnetic resonance (where
$\tomega=-m\omega_{A\phi}$) enters the flow, the wave absorptions at
the two magnetic resonances partially cancel, and the mode growth rate
starts to decrease again (see the right panels of
Fig.~\ref{fig:fl_omega}). In this case, the term ``corotation resonance'' simply refers to the combined effect of two magnetic resonances.

\subsection{Effects of different inner disc radii}

Here we consider the same setup as in Sections 4.1-4.2 except that the
inner disc boundary lies outside the ISCO.  The motivation for
considering $r_{\rm in} > r_{\rm ISCO}$ is that in real accreting
NS systems, the magnetosphere radius may well be outside the ISCO
(e.g., in accreting millisecond X-ray pulsars with surface magnetic 
field of $10^{8-9}$~G, the \Alfven radius is about 1.5-2 stellar
radii). Also, in the ``transitional state'' (when HFQPOs are observed) 
of BH X-ray binaries, the thin thermal disc may be truncated at 
a radius slightly larger than the ISCO (e.g., Done et al.~2007; 
Oda et al.~2010).

\begin{figure}
\begin{center}
$
\begin{array}{cc}
\includegraphics[width=0.47\textwidth]{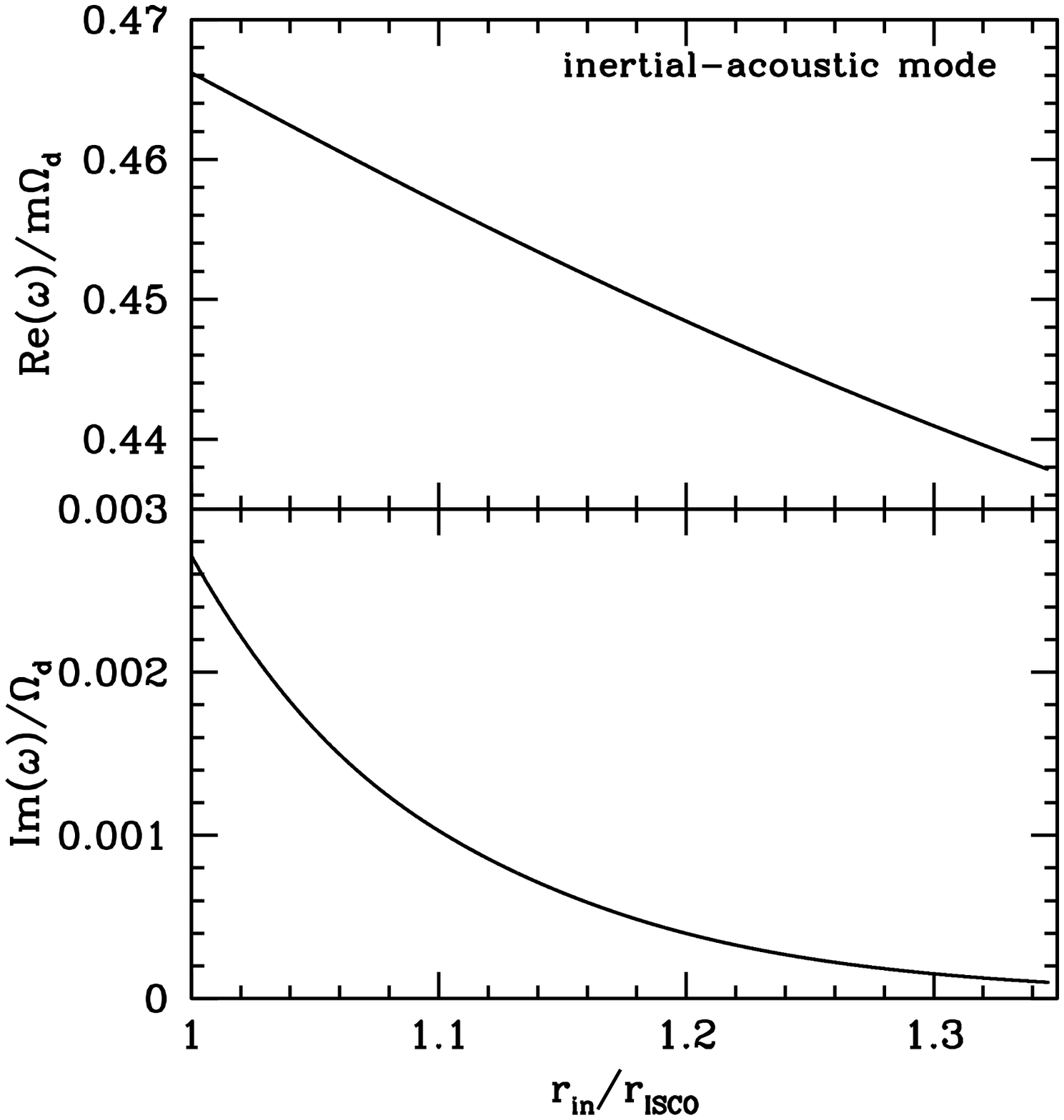} &
\includegraphics[width=0.47\textwidth]{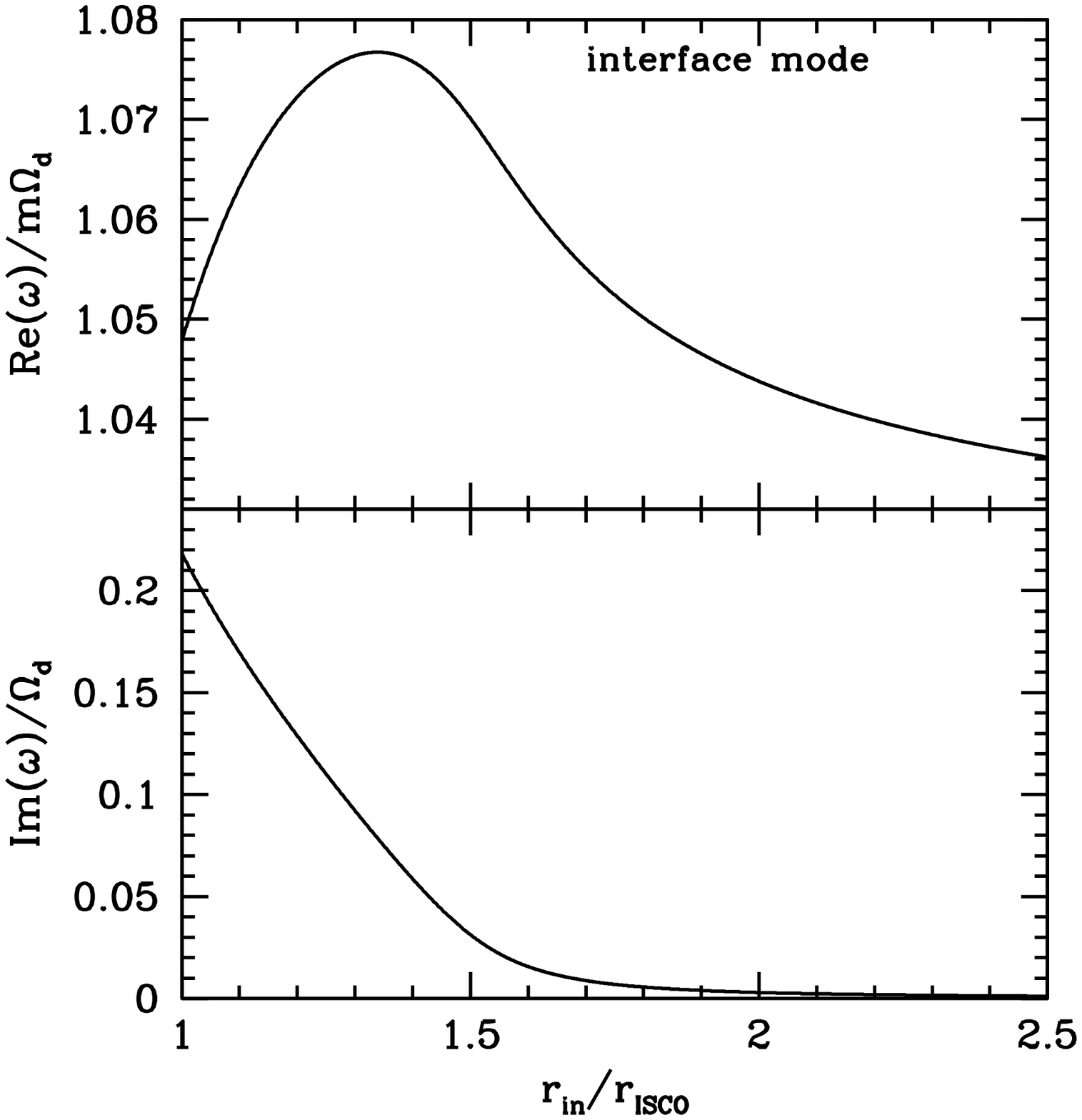}
\end{array}
$
\caption{The real and imaginary parts of the wave frequencies (in
  units of $\Omega_d$, the disc rotation rate at $r_{\rm in}$) for
  unstable inertial-acoustic modes (left panels) and unstable
  interface modes (right panels) as a function of the inner disc
  radius $r_{\rm in}$ (in units of $r_{\rm ISCO}$).  The magnetosphere
  inside $r_{\rm in}$ has zero density and the disc toroidal magnetic
  field is set to zero.  The other disc parameters are $m=2$,
  $c_{s}=0.1r\Omega$ and $\rho \propto r^{-1}$.}
\label{fig:omega_ib}
\end{center}
\end{figure}

\begin{figure}
\begin{center}
\includegraphics[width=0.55\textwidth]{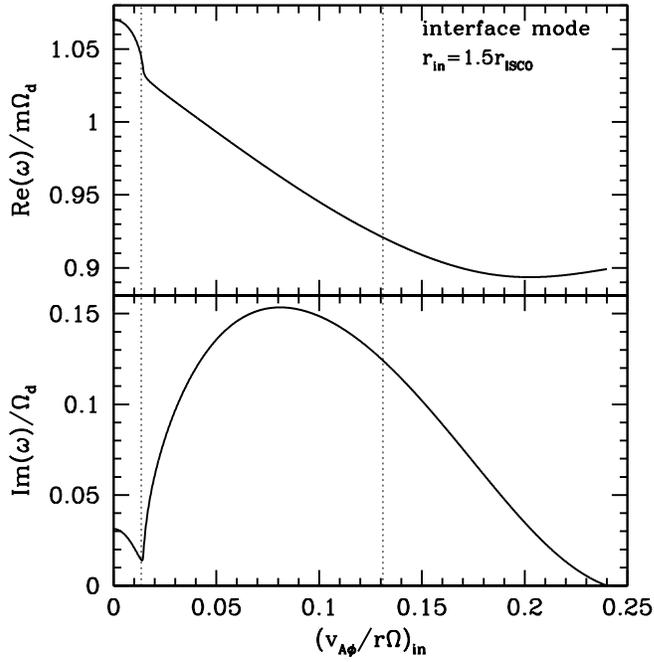}
\caption{Real and imaginary parts of the wave frequencies (in units of
  $\Omega_d$, the disc rotation rate at $r_{\rm in}$) for unstable
  interface modes as a function of the disc toroidal field strength,
  with the inner disc boundary located at $r_{\rm in}=1.5r_{\rm
    ISCO}$.  The $x$-axis specifies the ratio of the \Alfven velocity
  to disc rotation velocity at $r_{\rm in}$.  The magnetosphere
  density is set to zero.  The disc parameters are $r_{\rm in}=r_{\rm
    ISCO}$, $m=2$, $c_{s}=0.1r\Omega$ and $\rho \propto r^{-1}$. The
  two dotted vertical lines from left to right mark the entering of
  outer magnetic resonance and inner magnetic
  resonance into the flow, respectively.}
\label{fig:fl_omega2}
\end{center}
\end{figure}

As in Section 4.2, we assume that the magnetosphere has a negligible
density. Figure \ref{fig:omega_ib} shows an example (for a
non-magnetic disc) of how the (complex) mode frequencies depend on
$r_{\rm in}$.  When expressed in units of $\Omega_d$, the real
frequencies of both interface and inertial-acoustic modes are only
modestly affected by $r_{\rm in}/r_{\rm ISCO}$, but the growth rates
decrease rapidly with increasing $r_{\rm in}/r_{\rm ISCO}$. For the 
inertial-acoustic mode, this arises from the reduced vortensity slope
at distances further away from the ISCO, which results in 
smaller wave absorption at corotation. For the interface mode,
the larger vorticity just beyond $r_{\rm in}$ leads to a stronger 
rotational suppression of the RT instability, thus a smaller mode
growth rate.

Figure \ref{fig:fl_omega2} shows how the complex frequency of the
interface mode depends on the disc magnetic field $B_\phi$ when the
inner disc radius is set to $1.5r_{\rm ISCO}$. The behaviour of the
mode growth rate as a function of $(v_{A\phi}/r\Omega)_{\rm in}$ is
similar to the right panels of Fig.~\ref{fig:fl_omega}. Here, the mode
growth rate is small when $B_\phi=0$. So when the outer magnetic 
resonance enters the flow, wave absorption dramatically increases
the mode growth rate.

\section{Discussion and Conclusion}

In this paper we have studied the non-axisymmetric MHD modes and
instabilities in a 2D model of magnetosphere-disc systems (see
Fig.~1), as may be realized in accreting neutron star or black-hole
X-ray binaries (see Section 1). We have examined various physical
effects and parameters that can influence the global modes in the
system, including the density and magnetic field of the magnetosphere,
the velocity contrast across the magnetosphere-disc interface, the
rotation profile (Newtonian vs GR), the temperature and magnetic field of
the disc. We restrict to modes that do not have vertical structure, but
otherwise our calculations include all possible instabilities and
global oscillations associated with the interface and the disc. We
highlight several key findings and implications of this paper below.

\subsection{Interface instabilities in a rotating, magnetized system}

This paper includes a comprehensive study of the large-scale
Rayleigh-Taylor (RT) and Kelvin-Helmholtz (KH) instabilities
associated with the interface of a rotating, magnetized system.  
RT and KH instabilities have been studied intensively in
plane-parallel flows through both theoretical analysis and laboratory
experiments (e.g. Chandrasekhar 1961; Drazin \& Reid 1981), and have
found applications in various astrophysical and space environments.
But few papers have focused on rotating systems (e.g., Spruit et
al.~1995; Lovelace et al.~2009,~2010). Our study generalizes previous
works by Li \& Narayan (2004) and Tsang \& Lai (2009b) by considering
compressible fluid, magnetic field and rotation profile of the disc,
as well as generic field (poloidal and toroidal) configuration of the
magnetosphere.

As in plane-parallel flows, the interface modes are mainly driven
unstable by the RT and KH instabilities. Toroidal magnetic field
tends to suppress the instabilities through magnetic tension.
The magnetic field also indirectly affects the RT instability by modifying
the effective gravity. Except for the $m=1$ mode, for which the two opposite
effects of $B_\phi$ cancel (as conjectured by Li \& Narayan 2004), 
we find that increasing $B_\phi$ generally tends to reduce the growth rates of
the interface modes. Differential rotation (with finite vorticity) also 
tends to suppress the interface instability. To overcome this suppression
effect, the disc must have sufficiently large temperature (sound speed)
(see Tsang \& Lai 2009b). General relativity (GR) can significantly
affect the growth rate of interface modes because the disc rotation near 
the ISCO has smaller vorticity in GR than in Newtonian theory.

Another qualitatively new finding of this paper is that corotation
resonance can significantly influence the interface instabilities.  As
the toroidal field strength in the magnetosphere or in the disc
increases, the real frequency of the interface mode falls below
$m\Omega_d$ (where $\Omega_d$ is the disc rotation rate at the
interface), and corotation resonance (or its generalization to
magnetic resonances) appears in the disc. Wave absorption at
corotation can then significantly change the interface mode growth
rate (see Figs.~\ref{fig:pn_omega},~\ref{fig:fl_omega},~\ref{fig:fl_omega2}).

\subsection{Inertial-Acoustic Modes of relativistic Disc}
Our model system (Fig.~1) also accommodates inertial-acoustic modes
(p-modes) of relativistic discs. These modes are driven unstable
primarily by wave absorption at corotation resonance.  The
magnetosphere-disc interface naturally serves as the inner boundary
for the disc. Our study in this paper complements our previous works
(Lai \& Tsang 2009; Tsang \& Lai 2009c; Fu \& Lai 2011a) by properly
treating the inner boundary condition for disc oscillations.  Our
result shows that the magnetosphere behaves as a robust ``reflector''
for spiral waves in the disc: The p-mode frequency and growth rate do
not depend sensitively on the property (density, magnetic field) of
the magnetosphere. In agreement with Fu \& Lai (2011a), we find that
a modest disc toroidal field tends to reduce the growth rate 
of disc p-modes.

\subsection{Implications for High-Frequency QPOs in X-ray Binaries}
The large-scale ($m=1,\,2,\,3,\,\ldots$), overstable oscillation modes
studied in this paper may provide an explanation for the
high-frequency QPOs observed in NS and BH X-ray binaries (see Section
1). Obviously, the simplicity of our model setup precludes detailed
comparison with the phenomenology of QPOs.  But we note the following
relevant features of disc inertial-acoustic modes and interface modes.

(1) The disc inertial-acoustic modes have frequencies $\omega=\beta_d
m\Omega_d$ (with $\Omega_d$ the disc rotation rate at $r_{\rm in}$),
with $\beta_d<1$ (typically $\sim 0.5$) depending on model parameters
(such as disc sound speed) and boundary conditions. If the inner disk
is located at the ISCO, the mode frequencies can be computed {\it ab initio},
and the results are generally consistent with the observations of the
HFQPOs in black-hole X-ray binaries (see Lai \& Tsang 2009; Tsang \& Lai
2009c). The problem with these modes is that even a weak (sub-thermal)
disc toroidal magnetic field can suppress their instabilities
(see Section 4 and Fu \& Lai 2011a). Although large-scale
poloidal fields can enhance the instability under certain conditions
(see Tagger \& Pallet 1999; Tagger \& Varniere 2006), it is not
clear at this point which effects (disc toroidal field vs
large-scale poloidal field) will dominate.

(2) The interface modes have frequencies $\omega=\beta_i m\Omega_d$,
with $\beta_i\sim 1$. If $r_{\rm in}\simeq r_{\rm ISCO}$, the implied
QPO frequencies would be too high compared to observations. Of course,
as discussed before (see Sections 1 and 4.3), it is certainly possible
that $r_{\rm in}$ is somewhat larger than $r_{\rm ISCO}$ in real 
systems, but then we would lose the predictive power of our 
calculations (since in our model $r_{\rm in}$ is a free parameter).
On the other hand, the interface modes are robustly unstable, driven by
the RT and KH instabilities, and by the corotation effect, 
especially when the GR effect is included. In most cases, the growth rates
of the interface modes are much larger than the disc modes.

Overall, the results of this paper suggest that if the real accreting
NS or BH systems can be approximated by our magnetosphere-disc model,
the interface oscillations are more likely than inertial-acoustic oscillations to provide an explanation
for the observed QPOs. This is reasonable for NS systems, and is
consistent with spectral analysis of kHz QPOs in NS X-ray binaries
(e.g., Gilfanov et al.~2003). For BH systems, this would require that
the inner disc radius in the ``transitional state'' to be slightly
larger than $r_{\rm ISCO}$ (e.g., Done et al.~2007; Oda et al.~2010).

Nevertheless, it should be noted that our calculations of global disc modes are
based on a fairly primitive model. Several potentially important physical ingredients have been neglected in our setup, such as finite vertical wavenumber, disc turbulence due to MRI, radial mass inflow across ISCO, etc.. The effects of these factors on our results are certainly worth future investigations. Plus, the disc modes in our study are all linear modes which in principle become nonlinear eventually. Whether our results here are still valid during the nonlinear stage or not needs to be addressed by doing numerical simulations (Fu \& Lai 2012, in prep.). Moreover, we have not done any analysis regarding to how these global modes would manifest themselves observationally. In this sense, our study should better be treated as a demonstration of interesting physical mechanisms, rather than a mature theoretical tool for explaining or predicting HFQPO observations.

\section*{Acknowledgements}
We thank Prof. Wlodek Kluzniak for his careful reading of our manuscript and his valuable suggestions that helped improve our presentation. This work has been supported in part by NSF Grant AST-1008245 and NASA Grant NASA NNX10AP19G.  WF also acknowledges the support from a
Laboratory Directed Research and Development Program at LANL.


\label{lastpage}

\appendix
\section{General Magnetic Field profiles in the Magnetosphere}

In the main text of this paper we adopt a special toroidal magnetic field profile
$B_{\phi}\propto r$ in the magnetosphere so that the perturbation
equations have simple analytical solutions.
For a more general magnetic field profile $B_{\phi} \propto r^q$,
with $q>1$, analytical solutions are no longer attainable, and we must 
solve numerically the perturbation equations for 
both the disc (Eqs.~[\ref{eq:oded1}]-[\ref{eq:oded2}]) and the magnetosphere
(Eqs.~[\ref{eq:odem1}]-[\ref{eq:odem2}]). To this end, we first need to
derive the boundary condition at the the inner boundary of the
magnetosphere, which is close to the center of the system
($r=0$). This is done by requiring the solutions to be regular at small 
radius. As $r\rightarrow 0$, we observe that $\omega_{A\phi} \propto
B_{\phi}/r\rightarrow 0$. In this limit, Eqs.~(\ref{eq:odem1})-(\ref{eq:odem2})
then reduce to
\be 
\frac{d\xi_r}{dr}=-\frac{1}{r}\frac{\tomega-2m\Omega}{\tomega}\xi_r+\frac{m^2}{r^2}\delta h,
\ee
\be 
\frac{d\delta h}{dr}=(\tomega^2-r\Omega^2)\xi_r+\frac{2m\Omega}{r\tomega}\delta h,
\ee
i.e, recovering the hydrodynamic equations\footnote{See Fu \& Lai
  (2011c) for the derivation of the regularity condition for a even more
  general system (differentially rotating, magnetized,
  self-gravitating, etc.).}. These simplified equations can be readily
solved, leading to a relation between $\xi_r$ and $\delta h$:
\be 
\xi_r=\frac{m\delta h}{r\tomega(\tomega+2\Omega)}.
\ee
With this regularity condition implemented at some finite yet small
inner boundary $r_{\rm c}$, we carry out the integration in the
magnetosphere towards the interface, while at the same time we also
integrate the disc equations towards the same interface and employ
shooting procedure to solve for the eigenvalues by requiring solutions
from both regions satisfy the matching condition at the interface
(i.e., continuity of $\xi_r$ and $\Delta \Pi$). 
Note that the eigenvalues of the system now have two components: the wave
frequency $\omega$, and the relation between the solutions in
two regions, e.g., $\delta h(r_{\rm c})/\delta h(r_{\rm out})$ or
$\xi_r(r_{\rm c})/\xi_r(r_{\rm out})$.

\section{Plane-parallel flow with a compressible upper layer and a magnetized lower layer}

Consider a system with two separate fluid layers in a constant gravitational
field $\bb{g}=-g\hat{z}$. The upper layer ($z>0$) is a compressible fluid 
of density $\rho=\rho_{+}\mbox{e}^{-z/H_{z}}$ and constant horizontal velocity
$u_{+}$, and the lower layer ($z<0$) is an incompressible fluid of constant density
$\rho_{-}$ and horizontal velocity $u_{-}$. 
Here, $H_z=c_s^2/g$ is the density scale height with $c_s$ being the sound
speed, and both velocities are along the $x$-axis. 
This system was studied by Tsang \& Lai (2009b), but here we 
we add a uniform horizontal magnetic field $B_x$ in the lower
fluid. As in Tsang \& Lai (2009b) (correcting typos in their Appendix),
we apply perturbations of the form
$\mbox{e}^{{\rm i}kx-{\rm i}\omega t}$ to both layers and solve
for $\omega$ by demanding the Lagrangian displacement and Lagrangian
total pressure perturbation be continuous across the interface
between upper and lower fluids. Denoting
\[
\tomega_{+}=\omega-ku_{+},
\]
\[
k_z^2=(k^2-\tomega_{+}^2/c_s^2),
\]
\[
\tilde{k}=(\sqrt{1+4H_z^2k_z^2}-1)/2H_z,
\]
and
\[
\tilde{\rho}_{+}=\rho_{+}k/\tilde{k},
\]
the final solution for $\omega$ can be written as
\be 
\omega=\frac{k(\tilde{\rho}_{+}u_{+}+\rho_{-}u_{-})}{\tilde{\rho}_{+}+\rho_{-}}\pm\sqrt{-\frac{k^2(u_{+}-u_{-})^2\tilde{\rho}_{+}\rho_{-}}{(\tilde{\rho}_{+}+\rho_{-})^2}-\frac{kg(\rho_{+}-\rho_{-})}{\tilde{\rho}_{+}+\rho_{-}}+\frac{k^2B_{x}^2}{4\pi(\tilde{\rho}_{+}+\rho_{-})}}.
\label{eq:solution}
\ee
The three terms under the square root clearly have different physics
meanings. The first term signifies the Kelvin-Helmholtz instability
which disappears when velocity shear across the interface vanishes; 
the second term depicts the
Rayleigh-Taylor instability which only exists for non-zero density
contrast; the last term, the only difference between
Eq.~(\ref{eq:solution}) and Eq.~(A14) in Tsang \& Lai (2009b),
describes the stabilizing effect of the magnetic field that lies along
the direction of the perturbation wave vector. If we take the
incompressible and hydrodynamic limit (i.e., $c_s \rightarrow \infty$,
$H_z\rightarrow \infty$ so that $\tilde{k}\rightarrow k$,
$\tilde{\rho}_{+}\rightarrow \rho_{+}$, and $B_{x}\rightarrow 0$),
then the above equation reduces to Eq.~(B12) in Li \& Narayan (2004).

\end{document}